\documentclass{nature}
\usepackage{amsmath,amssymb,amsthm}
\usepackage[sort,super]{natbib}
\usepackage{setspace}
\usepackage{xcolor}
\usepackage{ulem}

\newcommand{\apj}{Astrophys. J.}

\usepackage{graphicx}
\makeatletter
\let\saved@includegraphics\includegraphics
\AtBeginDocument{\let\includegraphics\saved@includegraphics}
\renewenvironment*{figure}{\@float{figure}}{\end@float}
\makeatother

\usepackage[symbol]{footmisc}
\renewcommand{\thefootnote}{\fnsymbol{footnote}}

\begin{document}

\title{Solar System Formation Analog in the Ophiuchus Star-Forming Complex}

\author{John C. Forbes$^{1,}\footnote[1]{Corresponding author; jforbes@flatironinstitute.org}$\ , Jo\~ao Alves$^{2,3}$, Douglas N. C. Lin$^{4,5}$}

\maketitle

\begin{affiliations}
\item Center for Computational Astrophysics, Flatiron Institute, 162 5th Avenue, New York, NY, 10010, USA
\item University of Vienna, Dept. of Astrophysics, T\"urkenschanzstr 17, 1180 Vienna, Austria
\item Radcliffe Institute for Advanced Study, Harvard University, 10 Garden Street, Cambridge, MA 02138, USA
\item Department of Astronomy and Astrophysics, University of California, Santa Cruz, CA 95064, USA
\item Institute for Advanced Studies, Tsinghua University, Beijing 100086, P.R. China
\end{affiliations}

\setlength{\baselineskip}{4ex}

\begin{abstract}

Anomalies among the daughter nuclei of the extinct short-lived radionuclides (SLRs) in the calcium-aluminum-rich inclusions (CAIs)
 indicate that the Solar System must have been born near a source of the SLRs so that they could be incorporated before they decayed away
\citep{lee_demonstration_1976}. $\gamma$-rays from one such living SLR, $^{26}$Al, are detected in only a few nearby star-forming regions. Here we employ multi-wavelength 
observations to demonstrate that one such region, Ophiuchus, containing many pre-stellar cores that may serve as analogs 
for the emerging Solar System\citep{shu_star_1987}, is inundated with $^{26}$Al from the neighboring  Upper-Scorpius association \citep{diehl_radioactive_2010}, and so may provide concrete guidance for how SLR enrichment proceeded in the Solar System complementary to the meteoritics. 
We demonstrate via Bayesian forward modeling drawing on a wide range of observational and theoretical results that this $^{26}$Al likely 1) arises from supernova explosions, 2) arises from multiple stars, 3) has enriched the gas {\it prior} to the formation
of the cores, and 4) gives rise to a broad distribution of core enrichment spanning about two orders of magnitude. This means that if the spread in CAI ages is small, as it is in the Solar System, protoplanetary disks must suffer a global heating event.

\end{abstract}

\noindent
Multi-wavelength imaging of the Ophiuchus complex, from millimeter to $\gamma$-rays (see Fig. 1), unveils the interaction between the Ophiuchus clouds (greyscale) with a cloud of live $^{26}$Al (red), traced by its 1.8 MeV $\gamma$-ray emission\citep{pluschke_comptel_2001}. Despite the limited formal significance, about $2\sigma$, of the detection of Upper Sco from the COMPTEL map alone\citep{diehl_radioactive_2010}, the $^{26}$Al appears to be affected by the presence of the L1688 cloud, containing many well known prestellar dense-gas cores with disks and protostars\cite{Wilking2008-om} (the latter represented by the red dots in the bottom left panel). The total mass\citep{diehl_radioactive_2010, krause_surround_2018} of $^{26}$Al in this region detected with INTEGRAL is $M_{^{26} \mathrm {Al}} = (1.1 \pm 0.29) \times 10^{-4} M_\odot$, with comparable contributions from statistical and systematic uncertainty.

\noindent
The wind-blown appearance of the Ophiuchus
clouds (e.g., lower left panel) suggests a flow coming from the top right of the image, reinforcing a scenario where the $^{26}$Al 
umbrella around the Ophiuchus cloud is caused by the same flow originating from the Upper-Sco cluster in the Sco-Cen association\cite{Loren1989-cy}. Moreover, the Doppler shift detected in the $\gamma$-ray line\citep{krause_surround_2018} indicates that the hot, $^{26}$Al-enriched gas is moving towards the Sun. Since L1688 is between Upper Sco and the Sun\citep{zucker_large_2019} (see also the distance modeling in the methods), this reinforces a picture in which the $^{26}$Al-rich gas is impacting L1688, which may itself give rise to L1688's line-of-sight motion towards the Sun\citep{krause_surround_2018}. 

\begin{figure}
\centering
\includegraphics[width=0.75\linewidth]{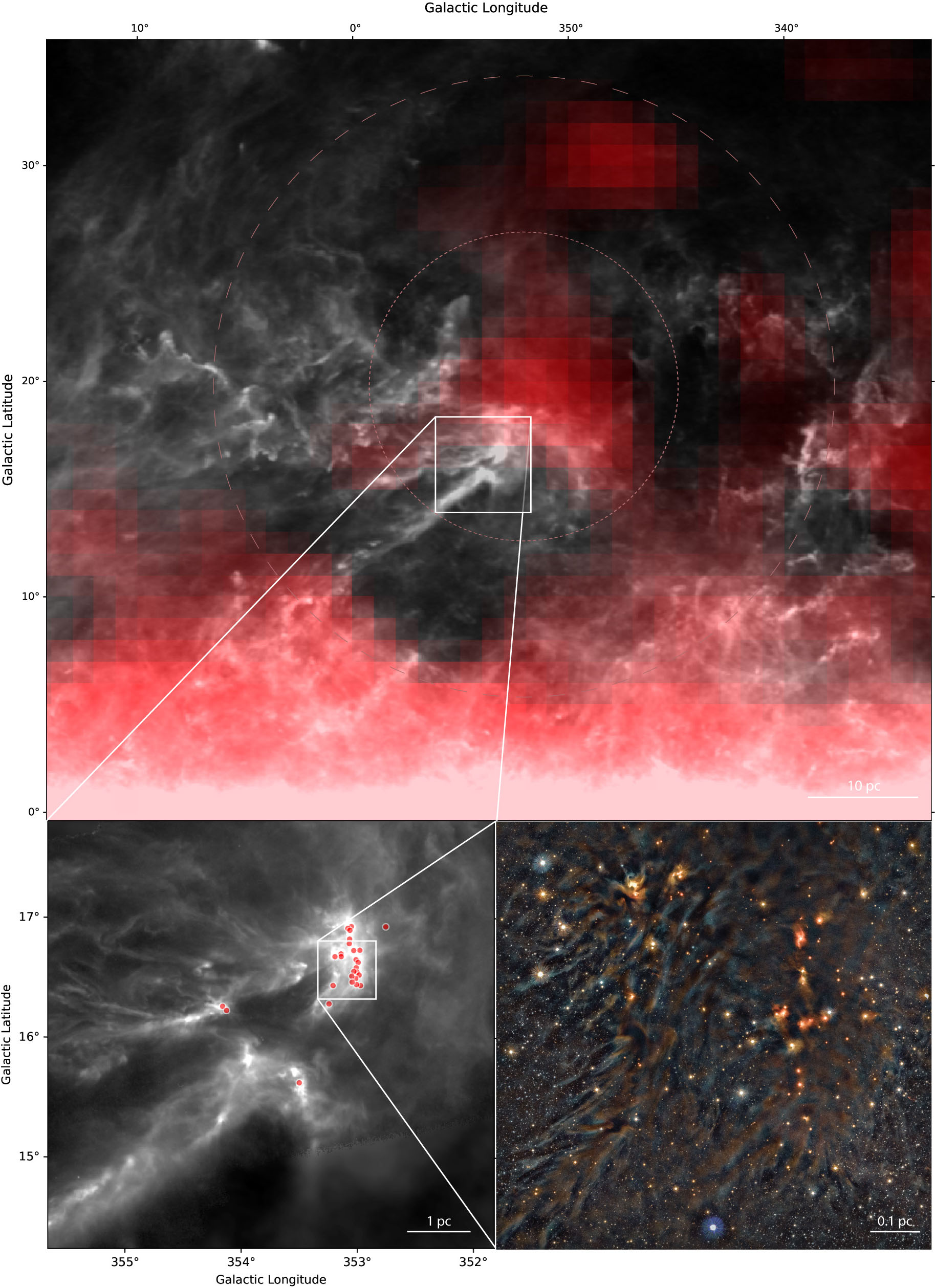}
\caption{
A multiwavelength view of Ophiuchus inundated by $^{26}$Al. Panel \textbf{a}) Distribution of $^{26}$Al (red) toward the Ophiuchus complex as measured by the COMPTEL satellite\cite{pluschke_comptel_2001}. The background grayscale represents column density derived from the Planck satellite's dust emission data \cite{abergel_planck_2014}. The dashed circles represent the 1$\sigma$ and 2$\sigma$ extent of massive stars in Upper Sco as estimated in the ``Distance distribution'' section of the Methods. Note that the cluster is at a larger distance than L1688, and that about 39\% (86\%) of massive stars should be within the projected area of the 1- (2-)$\sigma$ circle. The central box represents the area covered in panel \textbf{b}. Panel \textbf{b}) Distribution of Class I (protostars) in the Ophiuchus cloud\cite{kryukova_luminosity_2012} (red dots) displayed on a column-density map derived from Herschel data. The box centered on L1688 represents the area in panel \textbf{c}. Panel \textbf{c}) Deep NIR color composite of L1688 from the VISIONS ESO public survey, where blue, green and red are mapped to the NIR bands J (1.2 $\mu$m), H (1.6 $\mu$m), and Ks (2.2 $\mu$m) respectively. Selected dense cores and disks seen in scattered light are shown in Extended Data Figure 1. In all panels the scale bar assumes a distance of 140 pc from the Sun}
\label{fig:Fig1}
\end{figure}

\noindent
The Ophiuchus cloud complex contains many dense protostellar cores (Extended Data Fig. 1). 
With resolved sizes of the order of $\lesssim 10^4$AU, these irregular-shape cores are evidently able to persist despite, or perhaps because of, their close proximity to the massive stars in the Sco-Cen association. Class I protostellar objects inside these cores provide 
clear reassuring examples to alleviate the concerns on the initial and boundary 
conditions \cite{hester_hubble_1996, gounelle_origin_2008, mellema_dynamical_2006, gaidos_26al_2009, 
gritschneder_supernova_2012} raised in the context of different enrichment scenarios 
for the Solar System.

\noindent
While the presence of newly-forming stars so close to an abundant source of $^{26}$Al is suggestive, it is not obvious whether the $^{26}$Al comes primarily from supernovae (SNe) or Wolf-Rayet (WR) stars, how many individual sources are responsible, and whether and how the $^{26}$Al is likely to propagate to the dense cores in L1688, especially given that some or all of the relevant stars may already be extinct.
To address these questions, we construct a forward model for
the sources and history of $^{26}$Al production in the neighboring Upper-Sco young stellar 
association. In short, every single massive star that ever existed in the association is explicitly assigned a mass, birth time, and explosion flag, i.e. whether it would actually explode in a supernova or not. These quantities are treated probabilistically with carefully-constructed priors, taking into account the initial mass function (IMF), the mass of the association, literature estimates of the age of the association, and models of supernova explosions (see methods). In addition, the yields from both WR stars and SNe are allowed to vary with strongly-informed priors based on simulations\cite{sukhbold_core-collapse_2016, dwarkadas_triggered_2017} (see Extended Data Figures 2 and 3). The model is then constrained by the total mass of $^{26}$Al observed today in $\gamma$-rays by INTEGRAL. We obtain posterior samples and weights with nested sampling\cite{speagle_dynesty_2020}. With these samples, we can compute probable histories of Upper Sco, and with an additional set of assumptions as well as their attendant uncertainty, we can also quantitatively estimate different pathways for the incorporation of $^{26}$Al into the cores visible in L1688 (Fig. 2).

\noindent
Fig. 3 shows a key outcome of this modelling. Based on the weighted posterior samples, we can compute not only the probability that the present-day $^{26}$Al emission arises primarily from WR versus SNe, but we can also compute this probability conditional on the uncertain age of Upper Sco. In the figure, the opacity of the colors at each age represents the posterior probability that Upper Sco is that old -- that is, an age of 10 Myr is far more likely than an age of 13 Myr or 5 Myr. Integrating over all ages, a SN-dominated scenario (meaning that at least 90\% of the $^{26}$Al observed today is from SNe) occurs about 59\% of the time, whereas a WR-dominated scenario appears only about 27\% of the time (Fig. 2). This is an intuitive result, in that there are no currently-living WR stars in Upper Sco, but it is still possible for WR stars to have contributed more than 10\% of the currently-living $^{26}$Al, especially if the true age of Upper Sco is modestly younger than the current best estimate. Extended Data Fig. 4 shows a similar calculation for the number of individual sources responsible for 90\% of the $^{26}$Al, concluding that it is quite likely that more than one source was responsible for today's $^{26}$Al. Finally, we are also able to examine the probable histories and futures of $^{26}$Al levels in Upper Sco - Extended Data Fig. 5. shows examples of individual histories, making it clear that the mass of $^{26}$Al alive at any one time is highly variable, with contributions from a small handful of sources dominating since $^{26}$Al from previous events decays away rapidly. Extended Data Fig. 6 shows the median and central 68\% of the distribution at a series of times, demonstrating that on average the transition from WR-dominated to SN-dominated occurs around 7 Myr after the birth of Upper Sco.

\begin{figure}
\includegraphics[width=1\linewidth]{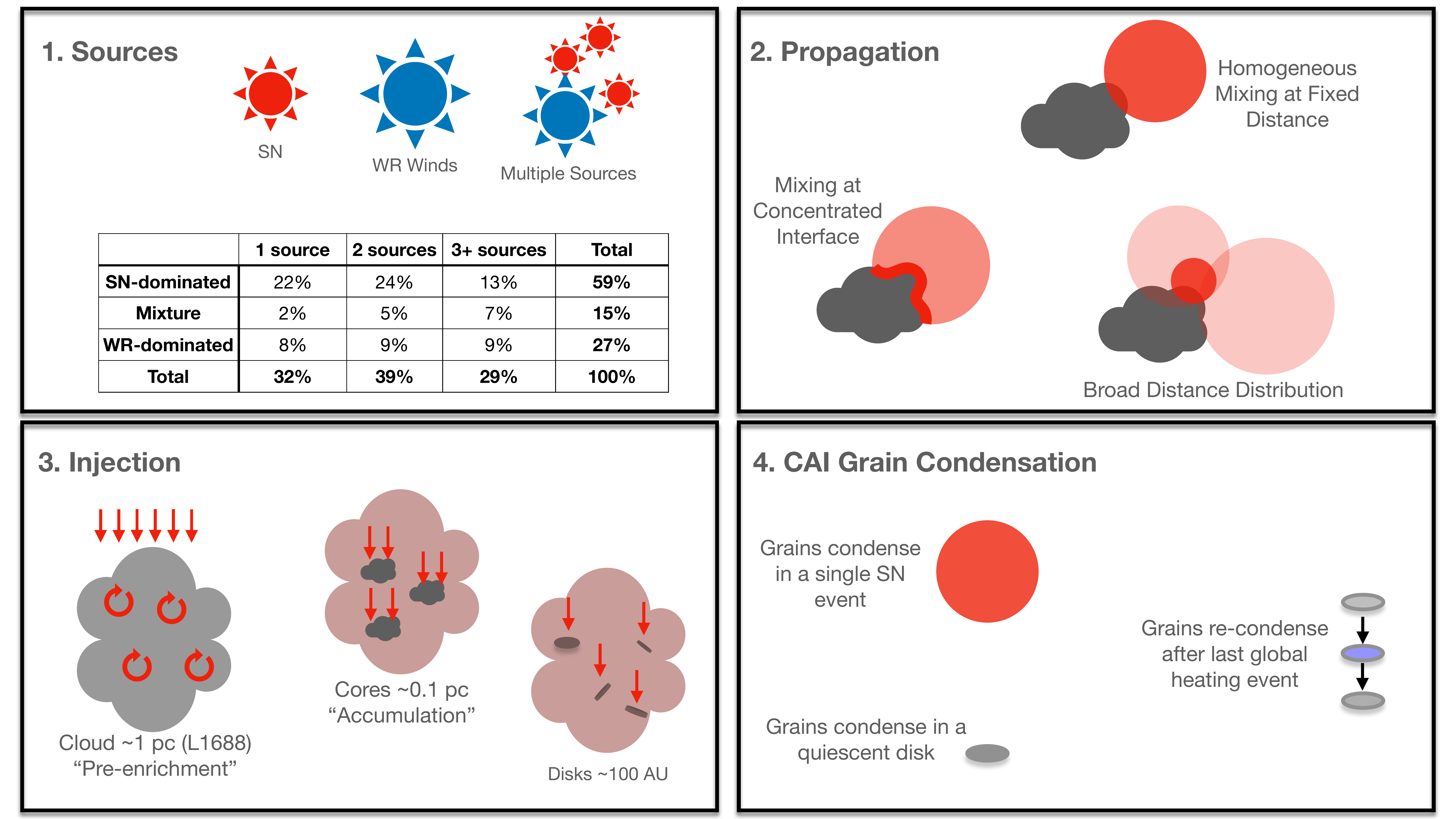} 
	\caption{Schematic of the steps involved from $^{26}$Al production through incorporation into CAI grains. The first panel summarizes our probabilistic modelling of the $^{26}$Al sources in Upper Sco, showing the joint posterior probabilities of the number and type of individual sources. Note that these probabilities are marginalized over all other variables in the model, most importantly the age of Upper Sco. The second panel shows several possibilities discussed in the text for how the $^{26}$Al propagates from the sources to L1688. The third panel shows possible routes for the $^{26}$Al to be incorporated into gas that will eventually form stellar systems, and finally the fourth panel shows different possibilities for the condensation of grains containing CAIs.}
\label{fig:cartoon}
\end{figure}

\begin{figure}
\centering
\includegraphics[width=0.9\linewidth]{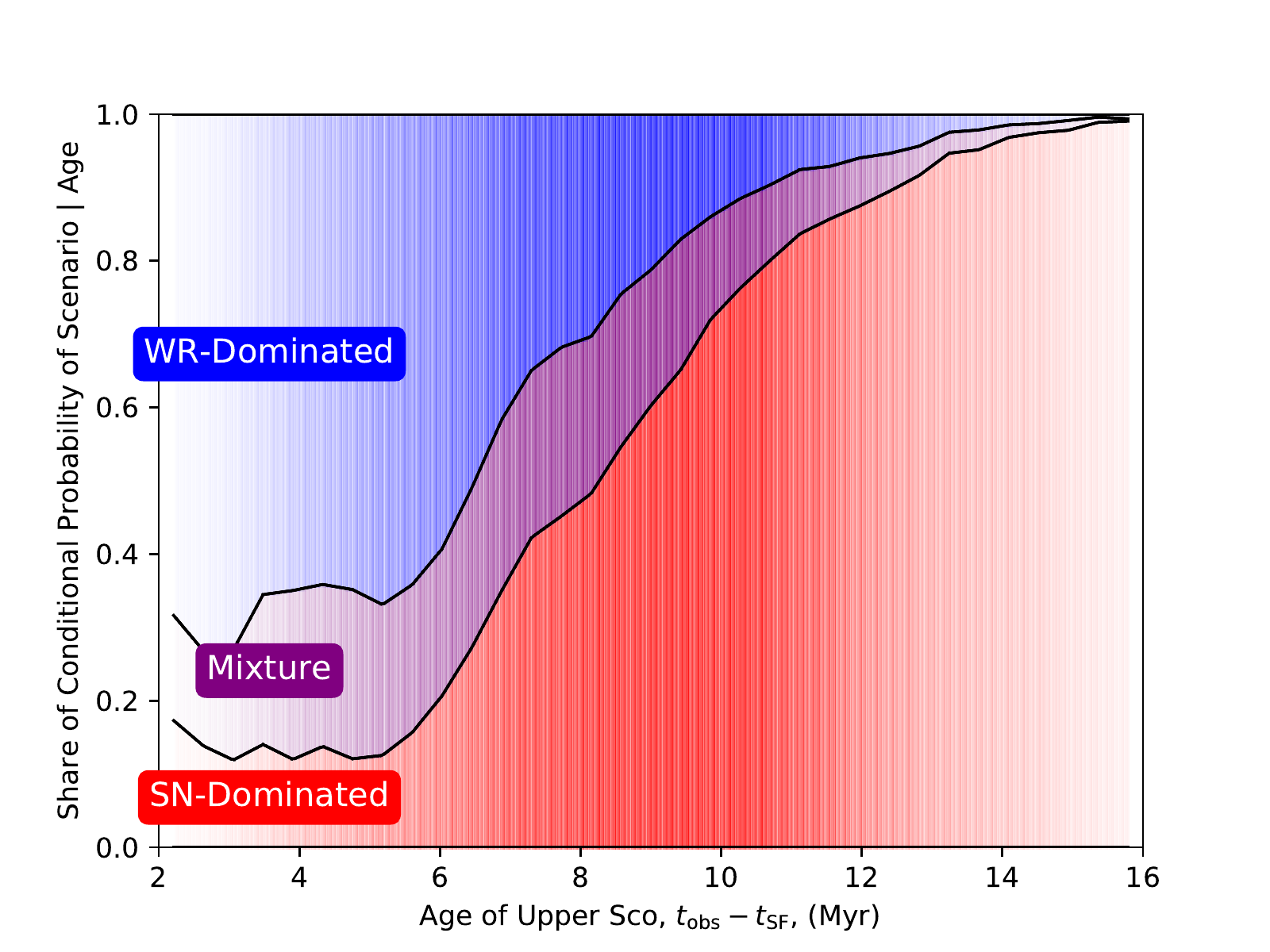} 
\caption{Origin of $^{26}$Al in Upper-Sco. Here we show the share of probability that the living $^{26}$Al observed in Upper-Sco comes predominantly from supernovae, Wolf-Rayet
stars, or a mixture of both, conditional on the average age of the association. The opacity shows the marginal posterior probability of a given age. The x axis refers to the time lapsed between the epoch of star formation and the current time. While we find that the supernova-dominated case is the most probable overall, a Wolf-Rayet dominated case is still possible, particularly if the association is modestly younger than that ($t_{\rm obs}-t_{\rm SF} \simeq 10$Myr) currently suggested by observations.}
\label{fig:SNvsWR}
\end{figure}

\noindent
Having fit an explicit model for the sources of $^{26}$Al in Upper Sco, we now turn to the question of how the $^{26}$Al may be incorporated into cores in Ophiuchus (see methods). Our model depends on an efficiency parameter $\epsilon$, the core radius $r_c$, the core lifetime $t_c$, the density contrast between the core and inter-core medium $\chi$, the relative velocity of these two media $v_\mathrm{rel}$, and a factor by which the $^{26}$Al is concentrated when it mixes into denser clouds $\mathcal{C}$. The results, shown in Fig. 4, we first draw the conclusion that regardless of the overall level of SLR enrichment, that enrichment is likely to be dominated by $^{26}$Al present when the core was formed, as opposed to $^{26}$Al accreted later. This is because $t_c \lesssim \tau_{1/2}$, i.e. the cores collapse so quickly\citep{enoch_mass_2008} that the radioactive aluminum only suffers a modest decay in the process. Meanwhile the accumulation term, at an order of magnitude level, is going to be smaller than the pre-enrichment term by a factor of $\sim \epsilon v_\mathrm{rel} t_c / ( \chi r_c ) \sim 0.01$ for our fiducial values, reflecting the fact that the core disappears before it can sweep up as much $^{26}$Al as it has access to at its formation.

\noindent
Next, we can compare to the Solar System's level of enrichment. For the range of plausible core ages\citep{enoch_mass_2008}, about $4.5 \times 10^5$ years, shown as the horizontal errorbar in Fig. 4, the pre-enrichment term on its own is sufficient to enrich cores to the Solar System's level most of the time, though there is a substantial scatter to take into account. Even with a fixed distance distribution and fixed values of the model parameters, the plausible (i.e. posterior) enrichment histories are such that the central 68\% of cases can easily span two orders of magnitude in $^{26}$Al mass accumulation. Additionally, $\mathcal{C}$ may vary substantially depending on whether the forming cores are within the region of compressive mixing. Interestingly the protostellar objects shown in panel b of Fig. 1 do fall along a well-defined ridge, which we speculate may correspond to a region of compressive mixing. We note, however, that the ridge does not face towards the source of SLRs, but rather towards ionizing radiation from the B star binary $\sigma$ Sco. Since the effects of ionizing radiation are not included in the simulations\citep{kuffmeier_tracking_2016, armillotta_mixing_2018} on which we base our estimate of $\mathcal{C}$, we acknowledge that the cores in L1688 may suffer substantial variation in $\mathcal{C}$.

\noindent
We therefore suggest that forming planetary systems have a bimodal distribution of SLR enrichment: the high-enrichment peak is broad, as just described, subject to varying distances from SLR sources and enrichment histories (as shown in Fig. 4), as well as variation in parameters that we can only estimate roughly, including the concentration factors $\mathcal{C}$, the density contrast $\chi$, the relative velocity $v_\mathrm{rel}$, and the core lifetime $t_c$. In addition to this broad peak, there is likely another set of stellar systems exposed to relatively little $^{26}$Al, some of which are visible in Fig. 1 in regions of Ophiuchus far from the current locus of SLRs. In the language of the propagation model, these are regions where $\mathcal{C}$ is much lower than our fiducial value of 100. Other star-forming complexes, e.g. Taurus, far from SLR sources, i.e. massive stars, likely fall entirely into this low-enrichment mode of the distribution\citep{reiter_observational_2020}. Note that enrichment to Solar System levels of $^{26}$Al does not fit neatly into any one idealized scenario discussed in the literature. The collapse of the cores in L1688 may be triggered by feedback from Upper-Sco, but not in a single blastwave\citep{gritschneder_supernova_2012,boss_triggering_2010}, and perhaps not in a well-defined shell around (possibly extinct) WR stars in Upper-Sco\citep{dwarkadas_triggered_2017}. It is not clear whether the gas in Ophiuchus was originally part of the same cloud that formed Sco-Cen and hence whether the continuous enrichment of a single cloud is the most appropriate idealized model\citep{young_inheritance_2014, vasileiadis_abundance_2013}, as opposed to the scenario\citep{fujimoto_short-lived_2018} in which the clouds are nearby on the scale of the Galaxy, and hence pollution from one enriches the other as it forms. The scenario we present here may be best thought of as a cloud-complex-scale version of the latter scenario, where the proximity of massive stars to future sites of star formation on $\sim 10$ pc scales is absolutely critical, and the SLRs themselves are not confined to a single GMC.

\begin{figure}
\includegraphics[width=1\linewidth]{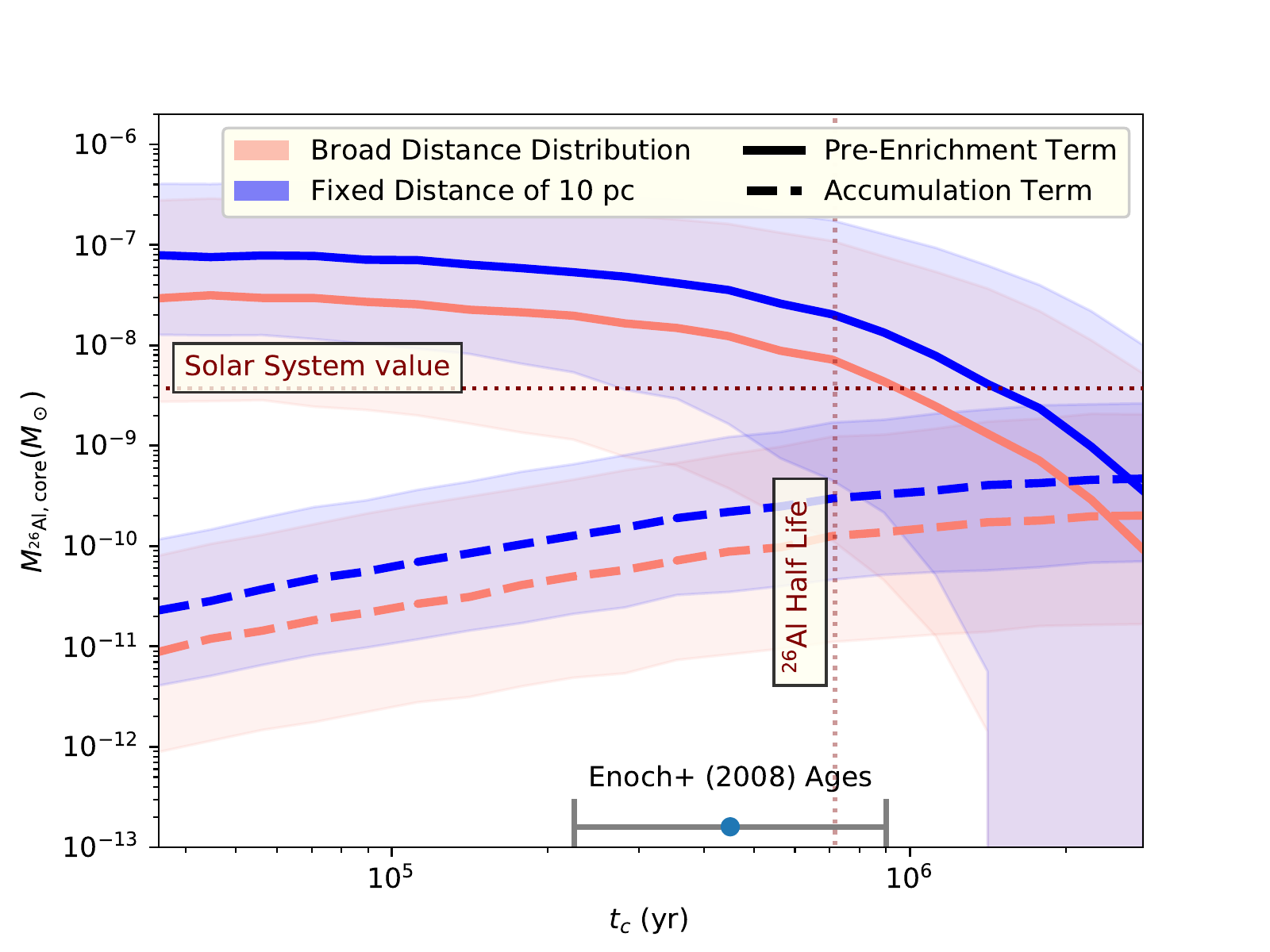} 
\caption{Enrichment of Cores. Here we show the total mass of $^{26}$Al incorporated into the present-day cores in L1688 under various assumptions. The lines and shaded regions show the median and central 68\% of the distribution over many draws from the posterior $^{26}$Al production model under different assumptions about the spatial distribution of sources and the different terms responsible for $^{26}$Al incorporation in the cores. The errorbar shows an observational estimate of core lifetimes in Ophiuchus\citep{enoch_mass_2008} and its standard error on the mean -- the cores likely collapse on a timescale shorter than the $^{26}$Al lifetime, meaning that most $^{26}$Al incorporated in cores arises from pre-enrichment rather than accumulation.   }
\label{fig:accum}
\end{figure}

\noindent 
\noindent 
Finally, we turn to the question of CAI grain formation.
The longevity of the protostellar cores and the multi-Myr persistence of the $^{26}$Al gas in the region (Extended Data Figs. 5 and 6) is in contrast to the narrow 
spread ($\lesssim 0.1$ Myr) in the formation episode of normal CAI's contained in the Solar System's CCMs \cite{jacobsen_26al-_2008, 
connelly_absolute_2012}. While the Solar System could be special in this regard, the Copernican Principle compels us to seek a means for the cores in Ophiuchus to obtain a similarly low age spread of CAIs. Since we expect that the vast majority of $^{26}$Al enrichment takes place via the ``Pre-enrichment'' rather than the ``Accumulation'' channel, there are two possibilities for grain formation that could give rise to narrow spread in lifetimes. First, the grains could form all at once prior to their incorporation in a core, i.e. in the remnant of a SN explosion. If they do not form all at once, e.g. if the $^{26}$Al comes from multiple sources, or the grains form gradually in a quiescent protostellar disk, then the grains must be heated and re-sublimated all at once over more or less the entire disk to reset and synchronize their radiogenic clock\cite{grossman_zoned_2002,
grossman_formation_2002, lodders_solar_2003}.  
One plausible mechanism for doing so is heating via episodes of rapid accretion, as in FU Ori objects\citep{hartmann_FU_1996}, though such outbursts may not be adequate to sublimate refractory grains and recondense CAIs at several AU. Another possibility is heating from a nearby supernova, $\lesssim 0.1$ pc from the protoplanetary disk\citep{portegies_zwart_consequences_2018, portegies_zwart_formation_2019}. We consider this to be unlikely given the distance distribution from massive stars (see Extended Data Fig. 7), though there is a young B-star, Elia 2-29, very near L1688. It is in the right place to serve this purpose for a small subset of the present-day cores, though not the majority given the $\sim 0.5$ pc radius of L1688, and it may be too young to explode by the time the present-day cores collapse and their disks dissipate. 
Yet another possibility is
via turbulent accretion that re-orients the angular momentum of the disk\citep{bate_chaotic_2010}, causing occasional catastrophic cancellation of angular momentum that shrinks the disk to a small size close to the parent star. Extended Data Fig. 1 shows that cores in L1688 have a head-tail, chaotic structure, unlike the comparatively symmetric, if somewhat elongated, cores in Taurus\citep{hacar_cores_2013}. We favor a global heating and re-sublimation scenario, since it is possible but unlikely that the $^{26}$Al in the region comes from a single supernova (see the first panel of Fig. 2).

\noindent 
Accretion of $^{26}$Al isotopes may continue onto protostellar disks after the core around them is depleted. However, the amount of live $^{26}$Al added during the class II protostellar disk evolution is small, and likely occurs after the reset of the radiogenic clock. For the most part, it is therefore reasonable to adopt the assumption that CAI's and chondrules had similar initial $^{26}$Al/$^{27}$Al ratios in the determination of their age differences ($\sim 2$ Myr) \cite{macpherson_calcium-aluminum-rich_2005, amelin_lead_2002, mckeegan_early_2003}.

\noindent
The common presence of CAIs in CCMs has long been considered evidence that SLRs were injected into the pre-solar cloud by nearby massive stars shortly before or during the collapse and buildup of the solar nebula\citep{lee_demonstration_1976, cameron_supernova_1977}. Based on the $\gamma$-ray emission of $^{26}$Al in the Ophiuchus complex, we show that this region of ongoing star formation, which may serve as an analog to Solar System formation, contains ample solar-mass protostellar cores. These cores can evidently not only survive in the proximity of the massive stars in Upper Sco, but we also show that they can be enriched in SLRs in amounts comparable to that found in the Solar System. The vast majority of the enrichment occurs prior to the formation of the core itself, a result found explicitly in simulations as well\citep{kuffmeier_tracking_2016}. Using the $^{26}$Al mass in the complex as the only constraint, along with priors on the cluster age, the initial mass function, and $^{26}$Al yields from SNe and WR, we find that SNe are more likely culprits for its production, though a WR origin 
cannot be ruled out. Multiple sources, rather than a single SN, are preferred. More stringent observational determination of the ages and masses of the stars in Upper-Sco would 
help to differentiate these scenarios.  
Since SNe generate $^{60}$Fe isotopes\cite{limongi_nucleosynthesis_2006, wang_spi_2007}
whereas WR stars do not, detection of $^{60}$Fe in $\gamma$-rays would also place constraints on the origin of SLRs\cite{gaidos_26al_2009} (see Extended Data Fig. 8). We expect that, while some systems will have very little $^{26}$Al enrichment, either because no massive stars have formed in the area, e.g. Taurus, or because they form in regions that do not compressively mix to $\mathcal{C} \sim 100$, many will have a level of enrichment comparable to the Solar System with a large scatter reflecting variability in mixing, enrichment histories, and distances from SLR sources. Given $^{26}$Al's central role as a heat source for CAI's, in the sublimation of volatile grains, and the coagulation 
of protoplanetary building blocks, this distribution is likely to be reflected in the population of exoplanets\citep{lichtenberg_water_2019, reiter_observational_2020}. Finally, we suggest that global heating and re-condensation event(s) are necessary to explain the small age spread in normal CAIs.

\begin{methods}
\subsection{Target Selection}

\renewcommand\thefigure{\arabic{figure}}    
\setcounter{figure}{0} 

\noindent
Young massive stars form in a small fraction of the volume of their parental giant molecular cloud. Still, their feedback has a profound effect on the rest of the molecular cloud, both compressing it and bringing it to collapse to form new stars (positive feedback) or eventually destroying it (negative feedback). Regardless of whether the massive stars end up in dense clusters or OB Associations,  configurations where massive stars interact radiatively and mechanically with dense gas soon after they are born is unavoidable.

\noindent
The Ophiuchus complex is the closest example to Earth of this common configuration of dense gas and massive stars. Star-forming gas in Ophiuchus is compressed by the feedback from a previous generation of massive stars (Upper-Sco, e.g., \cite{Loren1989-cy,loren_cobwebs_1989-1,De_Geus1989-my,De_Geus1991-sr,De_Geus1992-qc,krause_surround_2018}). Upper-Sco is a $\sim$10 Myr\cite{pecaut_star_2016, luhman_refining_2020,Sullivan2021-zx} subgroup in the Sco-Cen OB association with about 3500 young stars, including $\sim 20$ ionizing B stars and one O-star ($\zeta$ Oph). The massive stars and the Ophiuchus gas are intermingled and at about the same distance from Earth (Alves et al. 2021, in prep.). This proximity makes Ophiuchus a prime target, in terms of sensitivity and resolution, to examine questions of $^{26}$Al production and transport into dense cores and circumstellar disks.

\subsection{Data Used}
The $^{26}$Al data used in this Letter was observed by COMPTEL and taken from \cite{pluschke_comptel_2001}. The ESA Planck data used in the column-density map in Figure 1 was taken from\citep{abergel_planck_2014}. The ESA Herschel column density data for L1688 and the Ophiuchus streamers is taken from Alves et al. 2021, in preparation. This map was constructed following the data reduction procedure described in \cite{lombardi_herschel-planck_2014} that combines Herschel and Planck data calibrated with 2MASS infrared extinction maps. The NH$3$ data in Supplementary Info. Fig. 1 is taken from the GAS Survey\cite{Friesen2017-ob} and the $^{13}$CO from the COMPLETE survey\cite{Ridge2006-uz}. The NIR color-composite in Fig. 1 and Extended Data Fig. 2 use data taken from the ESO Public Survey VISIONS Data Release 1 (http://visions.univie.ac.at), a survey using the ESO VISTA telescope and the VIRCAM wide-field camera, and was reduced by Herv\'e Bouy and Stefan Meingast.

\begin{figure}
\includegraphics[width=1\linewidth]{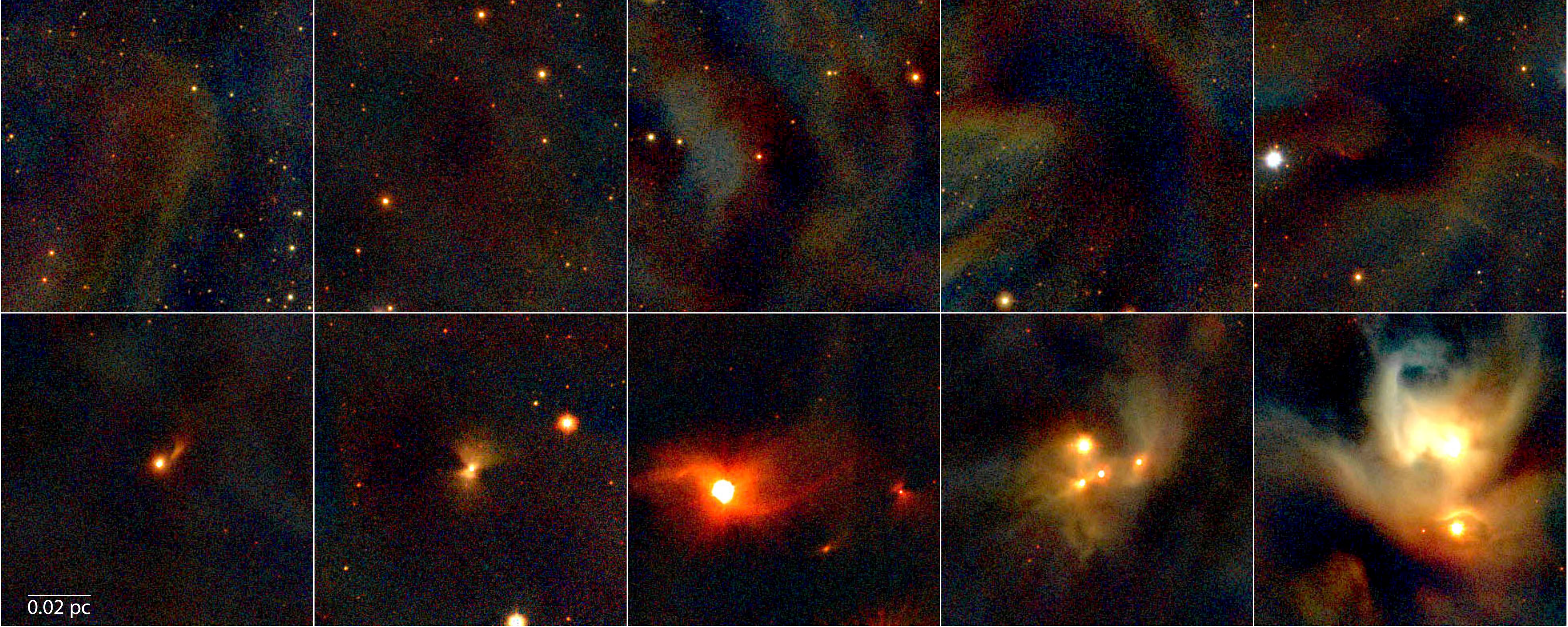} 
\caption{[Extended Data] Examples of cores and young stellar objects in L1688. The images are color composites from the VISIONS ESO public survey, where blue, green and red are mapped to the NIR bands J (1.2 $\mu$m), H (1.6 $\mu$m), and Ks (2.2 $\mu$m) respectively. In all panels the scale bar assumes a distance of 140 pc from the Sun}
\label{fig:cores}
\end{figure}

\subsection{Forward Modeling}

In order to better understand the origin of the $^{26}$Al observed by INTEGRAL \citep{diehl_radioactive_2010}, we present an explicit forward model for its production in the Upper Sco region. Ultimately our goal is to combine constraints on the age and mass of Upper Sco, along with constraints on yields from different $^{26}$Al sources to probabilistically determine when and by what sources the $^{26}$Al observed today was produced. This approach is particularly helpful both as a way to quantify uncertainty in the outcome, and just because we expect that some or all of these sources are no longer living today. The free parameters in this model are the mean age of the stars formed in this region, the 1-$\sigma$ (assumed Gaussian) spread of stellar ages, two parameters controlling the yields of $^{26}$Al produced by stellar winds, $A_{20}$ and $A_{120}$, and two parameters controlling the yields from supernovae, $f_\mathrm{SN}$ and $\zeta$. In addition to these population-level parameters, we model individual stars more massive than $8 M_\odot$ in this population. For each of these massive stars, we set its mass, its age relative to the mean age, and a Bernoulli (true-or-false) variable that denotes whether the star is destined to explode in a supernova. The number of stars, $N$ is fixed by the mass of the association, but the vast majority of these $N$ stars are less than $8 M_\odot$, and hence do not contribute to the production of $^{26}$Al on the timescales considered in this paper. This choice is made to keep the dimension of the inference problem the same regardless of the number of stars above $8 M_\odot$, while insuring that the a priori variance on the number of stars above $8 M_\odot$ is correct (see next subsection). In total this yields a set of $3 N+6$ unknown parameters, which we denote by the vector $\theta$. 

\subsection{Stellar Masses and Ages}
We impose a $10 \pm 3$ Myr normally-distributed prior on the cluster's mean age, and a flat prior from 0 to 3 Myr on the $1-\sigma$ age spread of the stars, based on current admittedly-uncertain observations\citep{pecaut_star_2016}. In other words, conditional on these two parameters, the prior of the age of the $i$th star is a gaussian with those two parameters. In addition to the priors mentioned above for the 4 yield-related parameters $f_\mathrm{SN}$, $\zeta$, $A_{20}$, and $A_{120}$, we impose a joint prior on the $N$ stellar masses such that $M_i>M_{i+1}$ and the prior on each $M_i$ is separately drawn from the IMF, i.e. $p(M) = \phi(M) \propto M^{-2.3}$ for $M>M_\mathrm{min}$. Note that imposing this ordering of the masses does not alter the functional form of the prior because for $N$ independent identically distributed random variables $X_i$, the corresponding sorted variables $X'_i$ have a joint distribution $p(X'_1,...X'_N) = N! \prod p(X_i)$.

\noindent
We now turn to the question of how to choose $N$ and $M_\mathrm{min}$. Since the dimensionality of the inference problem depends explicitly on $N$, it would be convenient to avoid varying it. Meanwhile the contribution to the $^{26}$Al mass in the region presumably only comes from stars with masses above 8 $M_\odot$. A priori, we do not know how many stars were born in the association with masses exceeding 8 $M_\odot$ -- even though some of these stars are around today, the models of young massive stars are reasonably uncertain, and of course some of these stars may have already died. To allow a variable number of stars to contribute to the $^{26}$Al production, we can choose a value of $M_\mathrm{min}<8 M_\odot$ so that some of the $N$ stars that we model are free to fall below 8 $M_\odot$. Clearly the lower $M_\mathrm{min}$ is, the more stars we will have to include in the model. However as $M_\mathrm{min}$ approaches $8\ M_\odot$, the model artificially restricts the number of stars with masses exceeding $8 M_\odot$ to exactly $N$. To find the appropriate balance, it is useful to think of the number of stars with masses exceeding $8\ M_\odot$ as a random variable with a binomial distribution, where the number of trials is $N$ and the probability of success is
\begin{equation}
    p = \frac{\int_{8 M_\odot}^\infty \phi(M) dM}{\int_{M_\mathrm{min}}^\infty \phi(M) dM}
\end{equation}
As $M_\mathrm{min} \rightarrow 0$, $p$ becomes small, of order a few percent depending on the exact IMF, and $N$ becomes large, so we could clearly approximate the binomial distribution in this limit as a Poisson distribution with expected value $N p$. The question is in practice, how small do we actually need to make $M_\mathrm{min}$ to retain this same Poisson distribution? We find that if we choose $M_\mathrm{min} = 2 M_\odot$, a Kroupa IMF\citep{kroupa_variation_2001} implies $p=0.165$ and observations of Upper Sco\citep{preibisch_exploring_2002} imply $N=76$. This implies that the number of stars above 8 $M_\odot$ will be $12.5 \pm 3.2$ per the binomial distribution, which is similar but does not have quite as large a variance as a more-accurate (but much higher-dimensional) model with a lower $M_\mathrm{min}$, which would have $12.5 \pm 3.5$ stars above $8 M_\odot$. For reference the cumulative mass function of the seven most-massive stars in Upper Sco is plotted along with posterior draws from the model's present-day mass function in Supplementary Data Fig. 2, demonstrating that the model, once fit to the present-day mass of $^{26}$Al, produces present-day mass functions in reasonable agreement with the observed mass function, even though the masses of individual massive stars are not used directly in the inference.

\subsection{ Supernovae} 
For a fixed value of $\theta$, we compute the total quantity of $^{26}$Al alive today by adding up the contributions from the supernovae and Wolf-Rayet phases of all $N$ stars. For the supernovae, we first check whether or not a given star has exploded -- the time since the star was born must be greater than the lifetime of the star, and the Bernoulli variable for that star, $S_i$, must be true. If these conditions are satisfied, the supernova contributes a mass of $^{26}$Al determined by linear interpolation of the log of the yield as a function of mass from tables of supernova yield and explosion models \citep{sukhbold_core-collapse_2016} - these values are of order $10^{-5} M_\odot$ per supernova, though they can be substantially higher for certain masses. Literature estimates for the $^{26}$Al yields vary by factors of a few at a given zero-age main sequence mass\citep{woosley_evolution_1995, rauscher_nucleosynthesis_2002, limongi_nucleosynthesis_2006, woosley_nucleosynthesis_2007}. The yields from each model in the literature have different mass dependencies, but they are not monotonic or otherwise easily-describable in a low-dimensional space. We therefore parameterize the uncertainty in a simplistic way, namely we multiply these supernova yields by a factor $f_\mathrm{SN}$, one of our population-level free parameters. We place a prior on $f_\mathrm{SN}$ that is normal in $\log_{10} f_\mathrm{SN}$, with mean 0 and 1-$\sigma$ width $\log_{10} 2$, i.e. a factor of two a priori uncertainty. 

\noindent
The prior on the values of the $S_i$ is more complex. The models\citep{sukhbold_core-collapse_2016} employ three different explosion mechanisms -- the Z9.6 model\citep{heger_nucleosynthesis_2010} for lower-mass stars, and both the W18\citep{sukhbold_core-collapse_2016} and N20\citep{nomoto_presupernova_1988, saio_nitrogen_1988, shigeyama_theoretical_1990} models for stars more massive than $\sim 12 M_\odot$. For a given mass, the yield of $^{26}$Al is similar between N20 and W18, but N20 produces systematically more explosions. Moreover, both explosion mechanisms exhibit complex behavior with mass, where in some mass ranges there are very few explosions, in others they are common, and in still others, explosions and non-explosions are intermixed, so that the explosion is best thought-of as a probabilistic event (see Extended Data Fig. 2). This is understood in the models as a result of the complex non-monotonic behavior of the stars' central density as a function of mass \citep{sukhbold_compactness_2014, sukhbold_core-collapse_2016}.

\noindent
To capture this behavior in our probabilistic model, we use a generalization of a logistic regression, where we assume that that probability of a supernova for a star of a given mass is
\begin{equation}
    \mathcal{S}_\zeta(M) = \frac1{1+\exp(f_\zeta(M))},
\end{equation}
where $f_\zeta(M)$ is the log-odds-ratio, and $\zeta$ is a parameter to be specified below. In an ordinary logistic regression $f_\zeta(M)$ would be a linear function, but to model the complex behavior seen in the supernova models, we use the more-general
\begin{equation}
    f_\zeta(M) = \sum_{j=1}^{N_j} C_{\zeta,j} \exp\left( -\frac12 \frac{(M-x_j)^2}{s_j^2}\right)
\end{equation}
We fix the $x_j$ to be the masses of every fifth explosion model run by Sukhbold et al, and the $s_j$ to be the separation between adjacent $x_j$. The goal is for this function to be extremely flexible, but to average over many different supernova explosion results to capture the seemingly probabilistic nature of the explosion process. We then fit the $C_{\zeta,j}$ by optimizing the following log-likelihood,
\begin{equation}
\label{eq:Lmech}
\ln L_\mathrm{mech} = \sum_k y_{\mathrm{mech},k} \ln \mathcal{S}_\zeta(M_k) + (1-y_{\mathrm{mech},k}) \ln ( 1- \mathcal{S}_\zeta(M_k))     
\end{equation}
where $k$ indexes the masses, $M_k$, of the models run by Sukhbold et al, and $y_{\mathrm{mech},k}=1$ if the model exploded for a given explosion mechanism $\mathrm{mech}$, and $y_{\mathrm{mech},k}=0$ if the model failed to explode.

\noindent
While this method allows us to specify the probability of an explosion as a function of mass, it is not obvious how to account for the uncertainty in the explosion mechanism. We proceed by defining the $\zeta=0$ case as the set of coefficients that optimizes the likelihood (Eq.~\eqref{eq:Lmech}) for the W18 mechanism, and the $\zeta=1$ case as the set that optimizes it for N20. We then define
\begin{equation}
    f_\zeta(M) =  f_{\zeta=0}(M) + \zeta (f_{\zeta=1}(M) - f_{\zeta=0}(M)).
\end{equation}
In other words, $f_\zeta(M)$ linearly interpolates between the W18 case ($\zeta=0$) and the N20 case ($\zeta=1$). The prior probability of a given explosion is then $P(S_i | M_i, \zeta) \equiv \mathcal{S}_\zeta(M_i)$. We also allow values of $\zeta$ outside this range, so our prior on $\zeta$ is uniform from -0.5 to 1.5, resulting in the prior on $S|M$ shown in Extended Data Fig. 2. By design, this prior exhibits the largest variation in the mass ranges where W18 and N20 produce different results.

\begin{figure}
\includegraphics[width=1\linewidth]{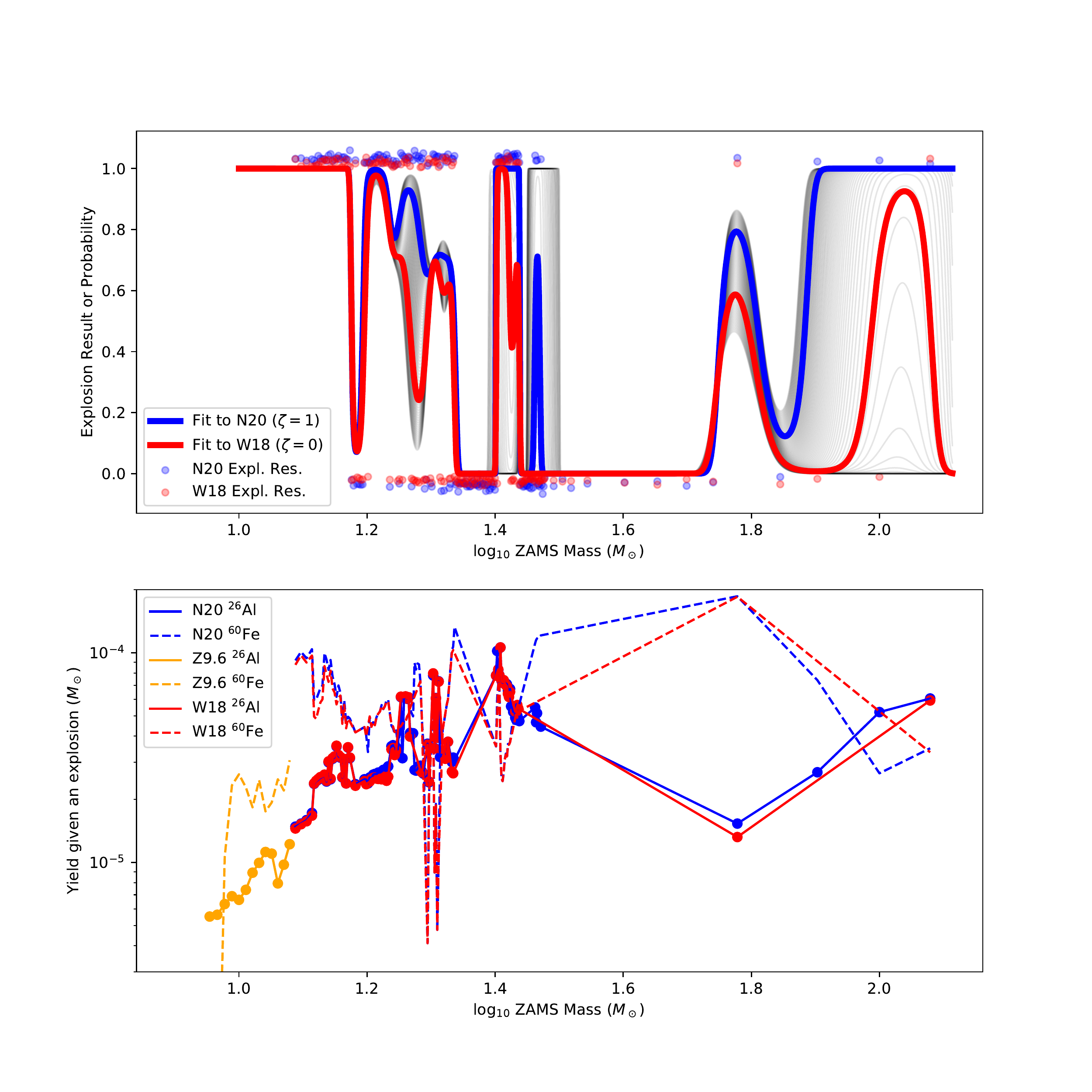} 
\caption{[Extended Data] The supernova prior, based on supernova explosion and yield calculations\citep{sukhbold_core-collapse_2016}. In the upper panel we show, as points near one or zero respectively, which models exploded or did not under different explosion mechanisms. We then fit a probability of explosion as a function of mass to these results (red and blue lines). The light grey lines show interpolation between (and extrapolation slightly beyond) these two scenarios, each of which represents our prior on whether a supernova of a given mass will actually explode conditioned on $\zeta$. The bottom panel shows the $^{26}$Al and $^{60}$Fe yields from these same calculations. We adopt the linear interpolation shown in the Figure of the N20 yields (the blue line) as the yields given an explosion has occurred. Between 8 and 12 M$_\odot$ we assume all stars explode and use the Z9.6 yields.}
\label{fig:Sprior}
\end{figure}

\begin{figure}
\centering
\includegraphics[width=0.8\linewidth]{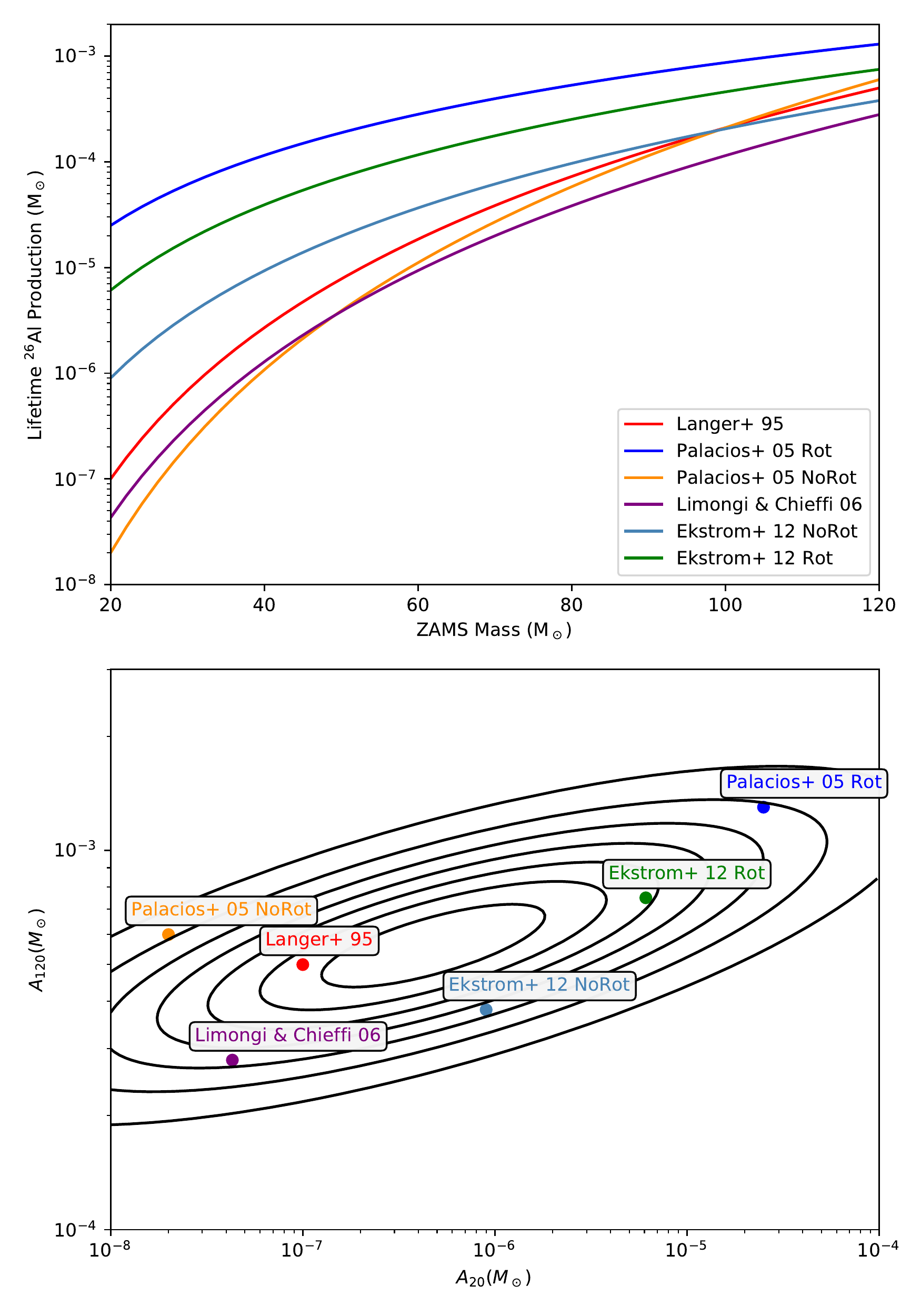} 
\caption{[Extended Data] The Wolf-Rayet prior. Following previous work\citep{dwarkadas_triggered_2017} we show the total lifetime production of $^{26}$Al from various yield calculations as a function of a star's mass from Langer\citep{langer_production_1995}, Palacios\citep{palacios_new_2005}, Limongi \& Chieffi\citep{limongi_nucleosynthesis_2006}, and Ekstr\"om\citep{ekstrom_grids_2012}. These are powerlaw fits to the actual simulation results. We then characterize these powerlaws by two numbers: the yield at 20 M$_\odot$ and the yield at 120 M$_\odot$, which we call respecitvely $A_{20}$ and $A_{120}$. Each study's result is plotted as a point in the lower panel. To assign a joint prior to the yields at these two masses, we fit a 2D gaussian to these points, essentially taking this collection of models to be a reasonable characterization of the possible values of $A_{20}$ and $A_{120}$.}
\label{fig:WRprior}
\end{figure}

\subsection{ Wolf-Rayet Winds} 
For the Wolf-Rayet phase, we use the set of theoretical models\citep{langer_production_1995, palacios_new_2005, limongi_nucleosynthesis_2006, ekstrom_grids_2012} compiled in previous work\citep{dwarkadas_triggered_2017}, and note that the total lifetime production of $^{26}$Al as a function of stellar mass in these models seems to be well-characterized by simple powerlaws, such that $M_{^{26}\mathrm{Al}} \approx a (M/M_\odot)^b$. These powerlaws could be characterized, instead of by $a$ and $b$, by the $^{26}$Al production at stellar masses of, say, $20 M_\odot$ and $120 M_\odot$, which we will call $A_{20}$ and $A_{120}$ respectively. {\it A priori} at $120 M_\odot$ there is only about an order of magnitude spread in the $^{26}$Al mass produced, but stars of $20 M_\odot$ have yields that vary by four orders of magnitude. It also seems that models that produce larger yields of $^{26}$Al at 120 M$_\odot$ tend to have, but are not guaranteed to have, larger yields for 20 M$_\odot$ stars. We therefore assume that, {\it a priori} $\log_{10} A_{20}$ and $\log_{10} A_{120}$ are jointly distributed normally with mean and covariance given by the sample of 6 models for Wolf-Rayet $^{26}$Al production (see Extended Data Fig. 3). We also assume that in all cases the yield from Wolf-Rayet stars is zero for stars with initial masses below $20 M_\odot$.

\noindent
In addition to the total mass of $^{26}$Al produced over the course of a star's Wolf-Rayet phase, we also need to account for the time-dependence of this production. Several models provide this information for the handful of different ZAMS stellar masses for which the calculation has been run. To avoid restricting ourselves to these masses, we rescale the production rate as follows
\begin{equation}
\frac{d M_{^{26}\mathrm{Al}}}{dt} = f_\mathcal{M} \left( t \times \frac{\mathrm{lifetime}_M}{\mathrm{lifetime}_\mathcal{M}} \right) \frac{M_{^{26}\mathrm{Al}, M, \mathrm{integrated}}}{M_{^{26}\mathrm{Al},\mathcal{M}, \mathrm{integrated}}}.
\end{equation}
Here $M$ refers to the mass of the star whose $^{26}$Al production we would like to estimate, and $\mathcal{M}$ refers to the mass of a model star where a given theoretical model explicitly prescribes the $^{26}$Al production rate, hence the $\mathcal{M}$ subscript in $f_\mathcal{M}$. For a given $M$, we choose the value of $\mathcal{M}$ closest in logarithmic space. We caution that the models do not behave monotonically, just as was the case with the supernovae, and using the nearest $f_\mathcal{M}$ rather than interpolating in the space of possible functions will lead to small discontinuities in these profiles as a function of mass. However, it is fair to say that in the models at a given rotation rate, each model's time-dependent $^{26}$Al production rate\citep{ekstrom_grids_2012} $f_\mathcal{M}(t)$ is broadly similar, with a large peak at $\sim 90\%$ of the star's lifetime. With rotation included, a substantial contribution from earlier in the stars' lives would appear as well; going forward we adopt the $f_\mathcal{M}$ with solar metallicity and zero rotation\citep{ekstrom_grids_2012}.
In other words, we assume that at a given fractional age, the star is producing a similar fraction of its lifetime $^{26}\mathrm{Al}$. All living $^{26}$Al is then attenuated by $2^{(t_\mathrm{prod}-t)/\tau_{1/2}}$, where $t_\mathrm{prod}$ is the time at which the $^{26}$Al was produced, and $\tau_{1/2} = 7.17 \times 10^5$ years is the half-life\citep{lee_demonstration_1976} of $^{26}$Al.

\subsection{Inference with the forward model}
\noindent
To fit the free parameters enumerated here, we use a dynamic nested sampler\citep{speagle_dynesty_2020} to obtain weighted samples from the posterior distribution $p(\theta|D) \propto p(\theta) p(D|\theta)$. Throughout the preceding paragraphs we have specified $p(\theta)$, the joint prior distribution on the model parameters. In addition to this, we need to specify the likelihood, and a transformation from the unit hypercube to the prior. The likelihood $p(D | \theta)$ is exceptionally simple: for a given value of $\theta$ we compute the quantity of $^{26}$Al alive today, which we denote $\hat{D}$. The likelihood is then a normal distribution in $\hat{D}$ centered at the observed\citep{diehl_radioactive_2010} value $D = M_{^{26}\mathrm{Al}} = 1.1\times 10^{-4} M_\odot$ with the observed $1\sigma$ uncertainty, $2.9 \times 10^{-5} M_\odot$, the quadrature-summed values of the quoted statistical and systematic uncertainties. 

\noindent
The transformation from the unit cube to the prior on $\theta$ is largely a matter of looking up the inverse CDF of the standard 1D distributions that specify independent components of the prior. However, one set of parameters requires some additional consideration. The masses of the $N$ stars could in principle be mapped from $N$ independent variables distributed uniformly from 0 to 1 by evaluating the inverse CDF of the IMF\citep{kroupa_variation_2001}, and assigning the most massive star to be $M_1$, the second-most massive to be $M_2$, etc. While this approach is technically correct, it will lead to a partition of the unit hyper-cube into $N!$ identical maps to the prior, which would be challenging for any nested sampling algorithm to handle. To avoid this, we instead compute the inverse CDF of the first order statistic, i.e. the maximum, then compute the inverse CDF of the second order statistic conditional on the first, then the third conditional on the second, and so on. Since the CDF of the maximum of $N$ independent identically distributed variables $X_1, ..., X_N$ is just $(F_{X_i}(x))^N$, we find that
\begin{equation}
    M_1 = M_\mathrm{min} (1 - u_1^{1/N})^{1/(1-\alpha)},
\end{equation}
where $u_i$ refers to the $i$th independent uniformly distributed variable to which we are mapping the prior, and $\alpha=2.3$ refers to the assumed slope of the IMF. For the $i$th order statistic conditional on the $(i-1)$st, we find
\begin{equation}
    M_i = M_\mathrm{min} \left[1 - u_i^{1/j}\left(1 - \left(\frac{M_{i-1}}{M_\mathrm{min}}\right)^{1-\alpha}\right)\right]^{1/(1-\alpha)}.
\end{equation}
This guarantees a one-to-one mapping between $u_1,...,u_N$ and $M_1,...,M_N$.

\begin{figure}
\includegraphics[width=1\linewidth]{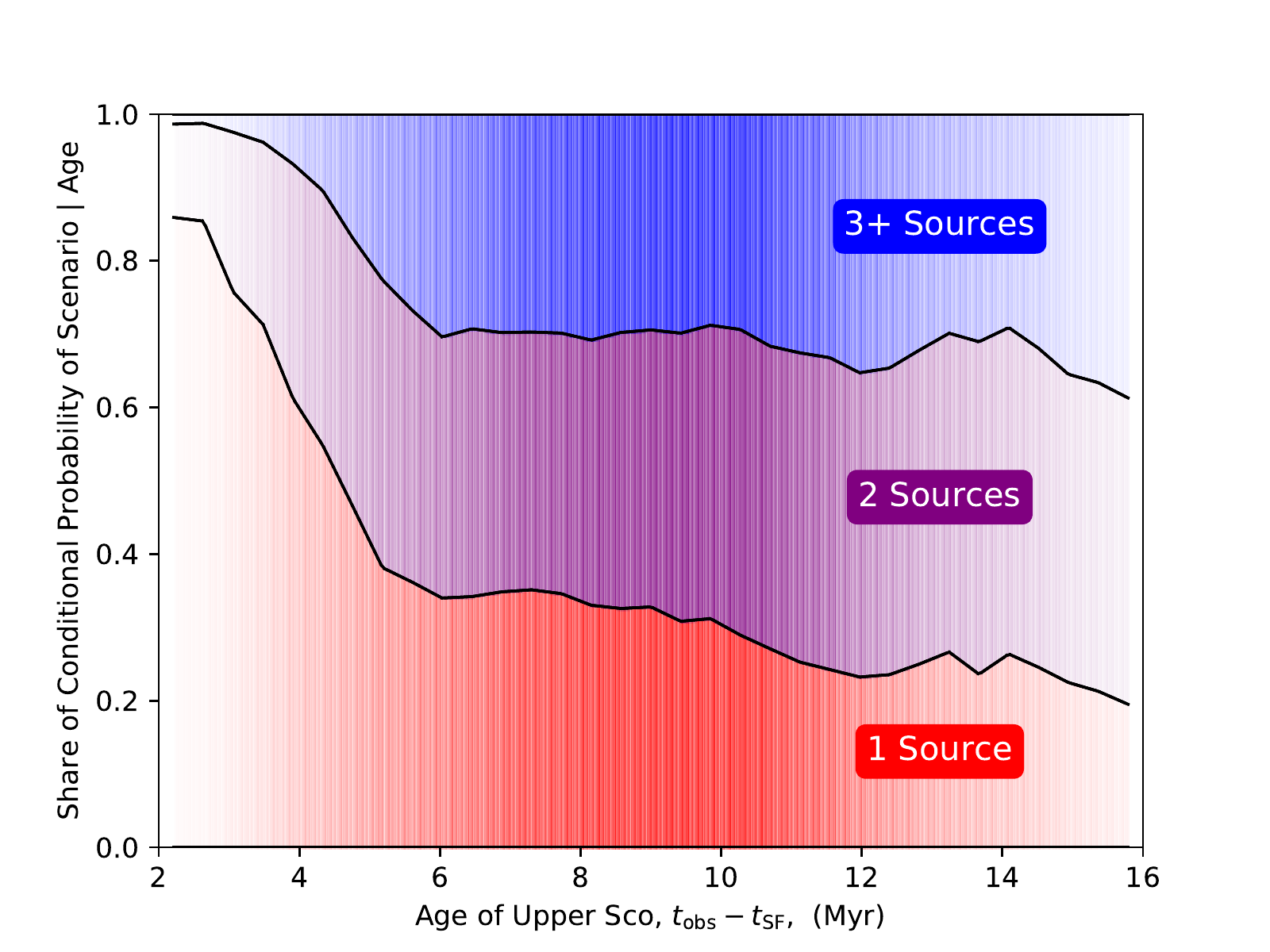} 
\caption{[Extended Data] The number of sources responsible for 90\% of the living $^{26}$Al in Upper-Sco. The opacity shows the marginal probability of each age, and at a given age, the fraction of the graph covered by each shaded region indicates the probability of that scenario. For all plausible values of the age of Upper-Sco, the living $^{26}$Al is likely to come from just a small handful of individual massive stars, but likely more than one.}
\label{fig:nsources}
\end{figure}

\begin{figure}
\includegraphics[width=1\linewidth]{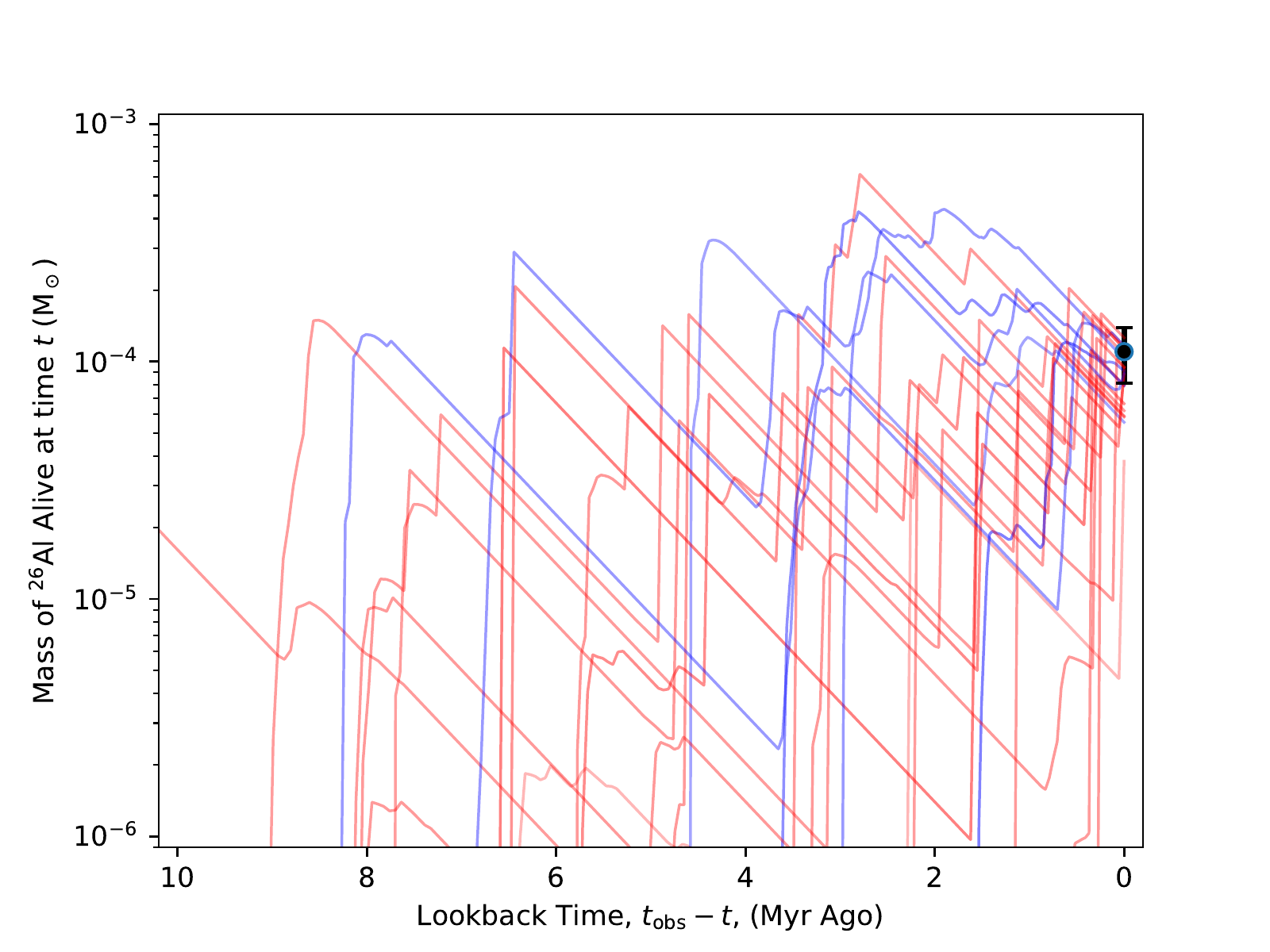} 
\caption{[Extended Data] Individual realizations of the mass of $^{26}$Al over time. The present-day observed value\citep{diehl_radioactive_2010} is shown as a point with errorbars representing the standard error on the mean on the right side of the plot. Each line is a different posterior realization of Upper-Sco, with transparency proportional to its weight. The colors show different scenarios. Blue lines have their $^{26}$Al today dominated by Wolf-Rayet stars, and red lines are dominated by Supernovae.  The characteristic sawtooth pattern of the lines comes about as individual sources of $^{26}$Al produce the radioactive isotope in a relatively short period of time (or instantaneously for supernovae), followed by exponential decay, as expected for young star-forming regions\citep{young_inheritance_2014}}.
\label{fig:alRealizations}
\end{figure}

\begin{figure}
\includegraphics[width=1\linewidth]{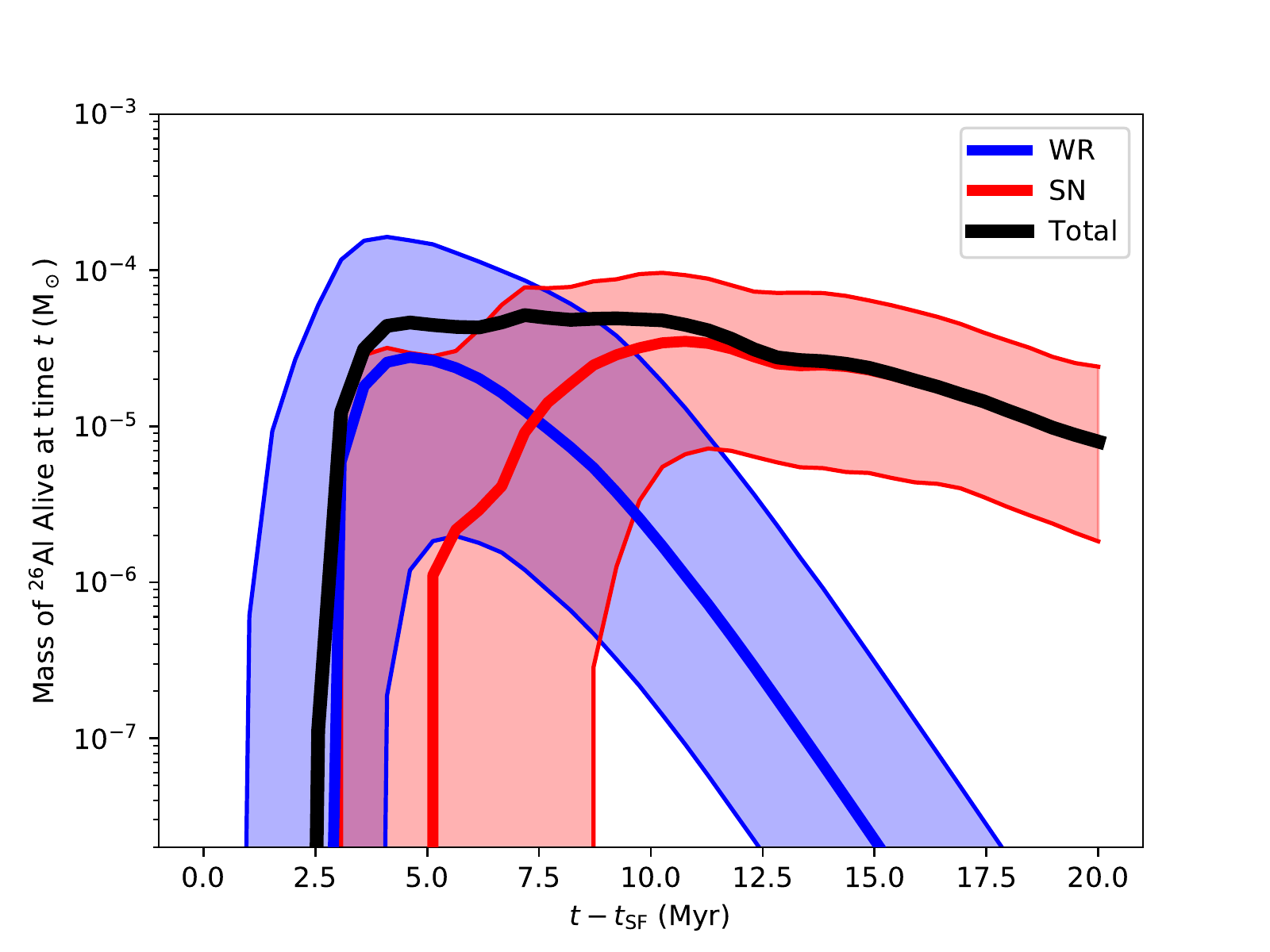} 
\caption{[Extended Data] The mass of living $^{26}$Al as a function of time since the birth of Upper-Sco. The blue shaded region shows the contribution from Wolf Rayet winds, and the red shaded region shows the contribution from supernovae. The thick lines show the median, and the regions show the central 68\% of the posterior contribution from each source at that time. Upper-Sco itself is at an age where the dominant contribution is shifting from Wolf Rayet winds to supernovae. The two conspire to generate a roughly constant quantity of living $^{26}$Al as a function of age in an average sense, though individual realizations of the cluster oscillate substantially as individual sources produce their $^{26}$Al, which then decays away until the next $^{26}$Al-producing event (see Extended Data Fig. 5).}
\label{fig:alDistr}
\end{figure}

\begin{figure}
\includegraphics[width=1\linewidth]{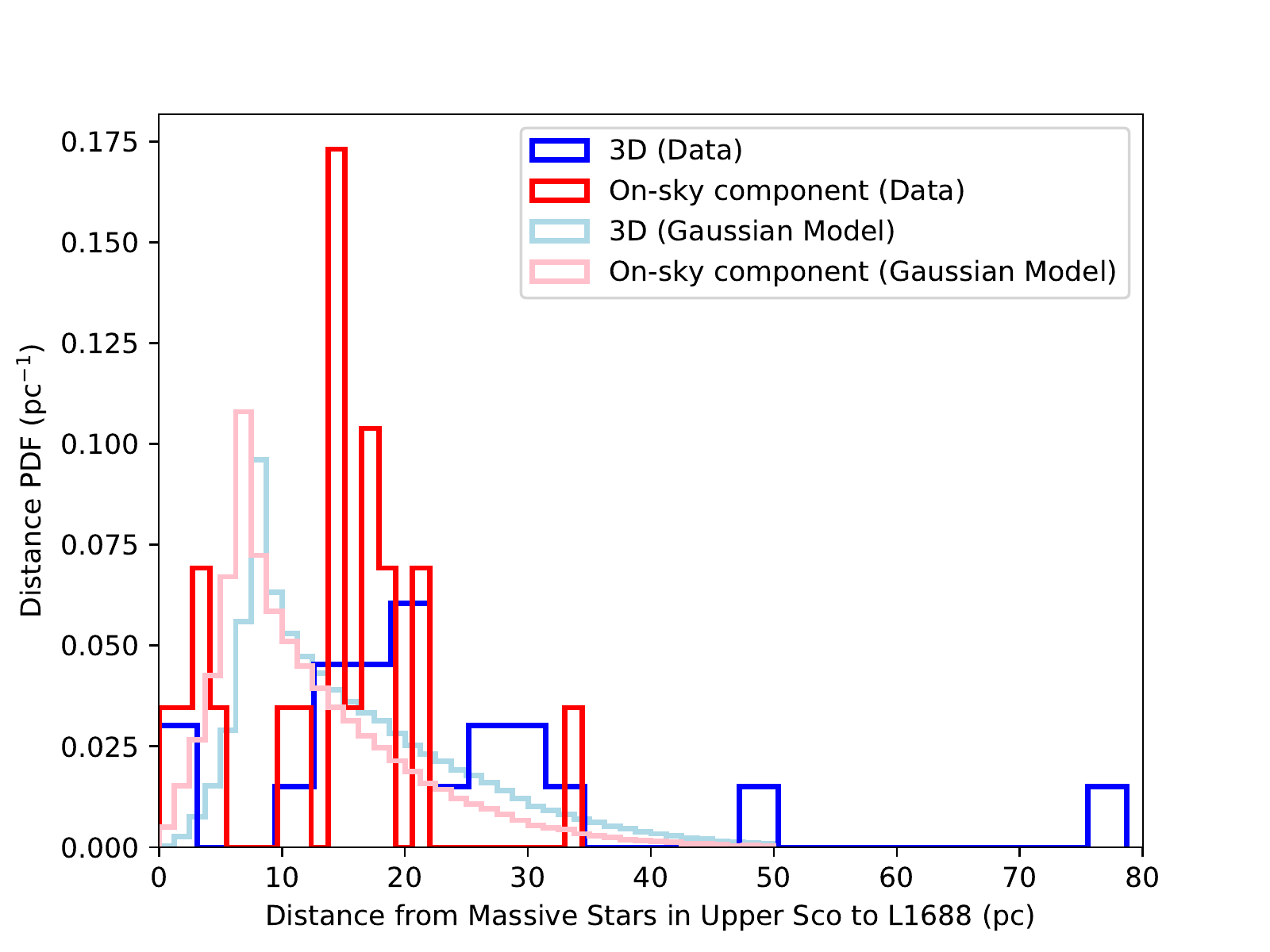}
\caption{[Extended Data] The distance distribution between L1688 and massive stars in Upper Sco. The data for 21 stars are shown in blue and red (the 3D distances and their on-sky component at the fiducial distance of L1688, respectively). In light blue and light red, we show the maximum {\it a posteriori} estimate for a single-component 3D gaussian model taking into account distance uncertainties on the individual massive stars.}
\label{fig:distdist}
\end{figure}

\begin{figure}
\centering
\includegraphics[width=0.8\linewidth]{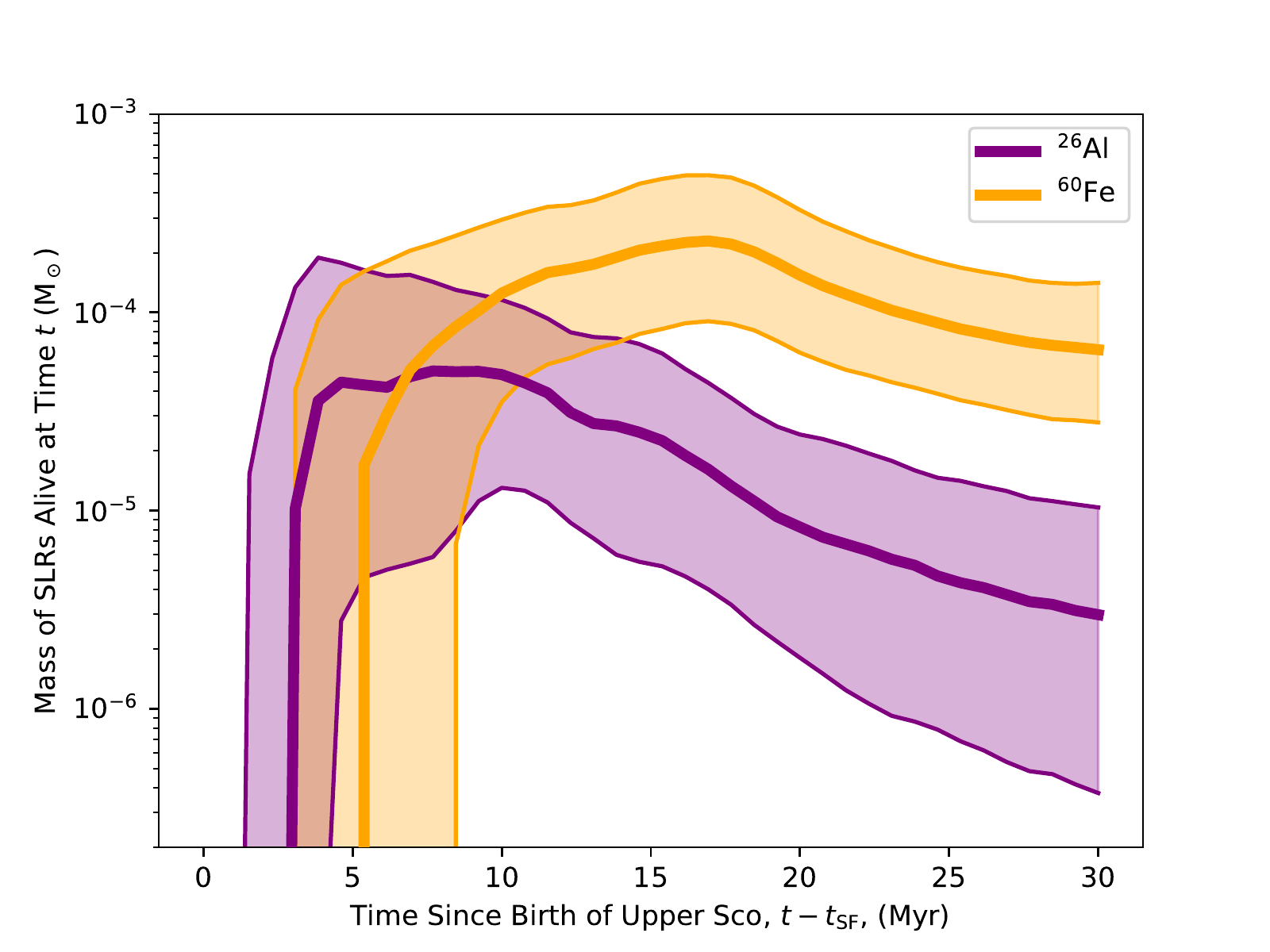} 
\caption{[Extended Data] Evolution of $^{60}$Fe compared to $^{26}$Al. The shaded area shows the central 68\% of the posterior distribution at that time (i.e. this distribution is a prediction for Upper Sco in particular). $^{60}$Fe peaks later than $^{26}$Al both because of its longer half-life and the lack of an early contribution from Wolf-Rayet winds.}
\label{fig:iron}
\end{figure}

\subsection{Comparison to literature results}
\noindent
 Assuming that the $^{26}$Al can be incorporated into cores and CAI grains, as discussed below, it is worth briefly mentioning how this result compares to past modelling. Initial stellar population synthesis of SLRs in Upper Sco\citep{voss_using_2009} concluded that WR stars were likely responsible for the observed $^{26}$Al, though, as is common, they adopt a particular set of stellar models and yields. Other recent studies \citep{gounelle_solar_2012, young_bayes_2016, dwarkadas_triggered_2017} also favor strongly WR-influenced scenarios, in part because of a meteoritic abundance of $^{60}$Fe that was revised downwards\citep{tang_abundance_2012} and in part because recent supernova nucleosynthesis models\citep{sukhbold_core-collapse_2016} suggest that stars above $\sim 30 M_\odot$ have difficulty exploding as supernovae (see Extended Data Fig. 2), leaving more room for the WR scenario\citep{young_inheritance_2014, young_bayes_2016, young_birth_2020}. Our present study focuses almost exclusively on $^{26}$Al because it is observable in the $\gamma$-rays, whereas $^{60}$Fe has so far not been convincingly detected in any particular region\citep{wang_gamma-rays_2020}. Purely based on the $\gamma$-rays and priors on the model parameters as discussed in the Methods, we favor SN sources, but can not rule out a WR source, particularly if Upper Sco is modestly younger than 10 Myr. 

\subsection{Enrichment Models}
\noindent
We now turn to the question of how and whether the cores in the Ophiuchus complex can attain a level 
of $^{26}$Al enrichment comparable to that contained in the Solar System CCMs\citep{lodders_solar_2003}. Given the short lifetime of $^{26}$Al the SLRs need to be incorporated quickly\citep{amelin_lead_2002, macpherson_calcium-aluminum-rich_2005, villeneuve_homogeneous_2009}. Potential SLR entry ways include 1) the continuous pollution of a parent GMC within which the protosolar cloud  
emerged\cite{arnould_short-lived_1997, gaidos_26al_2009, tatischeff_runaway_2010, fujimoto_short-lived_2018}, 
2) the accretion onto and/or the triggered implosion of the protosolar core \cite{cameron_supernova_1977, 
huss_stellar_2009, boss_triggering_2010}, and 3) the interception of the SLR-bearing gas or grains by the 
primordial solar nebula \cite{gounelle_origin_2008, adams_birth_2010, ouellette_injection_2010}. Accretion of $^{26}$Al isotopes may continue onto protostellar disks after the core around them is depleted. However, the amount of live $^{26}$Al added during the class II protostellar disk evolution is small, and likely occurs after the reset of the radiogenic clock. For the most part, it is therefore reasonable to adopt the assumption that CAI's and chondrules had similar initial $^{26}$Al/$^{27}$Al ratios in the determination of their age differences ($\sim 2$ Myr) \cite{macpherson_calcium-aluminum-rich_2005, amelin_lead_2002, mckeegan_early_2003}. We note that if the cloud as a whole were able to mix perfectly, the $^{26}$Al/$^{27}$Al ratio in Upper Sco and many other star-forming regions would be comparable to the Solar System's level of enrichment\citep{jura_26al_2013, young_inheritance_2014, young_bayes_2016, young_birth_2020, reiter_observational_2020}, and indeed these same models are able to explain many meteoritic abundances simultaneously. However, it is far from clear that mixing on this scale is possible. We therefore attempt to model the propagation of $^{26}$Al from Upper-Sco to Ophiuchus explicitly, taking advantage of the observational knowledge of the geometry of this particular region.

\noindent
We consider two possibilities: 1) the cores form from gas that has already been enriched with $^{26}$Al, and 2) the cores accumulate $^{26}$Al as enriched, lower-density, gas flows past (see the third panel of Figure \ref{fig:cartoon}). For the sake of simplicity, we assume that cores live for a fixed time $t_c$ with a fixed size $r_c$, and that the $^{26}$Al continues its decay until the end of the core lifetime, at which point its relative abundance is frozen in to the forming grains. We refer to the first case as the ``Pre-Enrichment'' term, and to the second as the ``Accumulation'' term. For both terms, we assume that the lower-density medium surrounding the cores has a density of $^{26}$Al, $\rho_{^{26}\mathrm{Al}}$, equal to $3 \mathcal{C} \sum_i M_{^{26}Al,i} / 4\pi \mathcal{R}_i^3$, namely a constant $\mathcal{C}$ representing the effects of small-scale mixing times the mass of $^{26}$Al from each of the individual sources in Upper Sco, $M_{^{26}\mathrm{Al},i}$ divided by the volume of a spherical region of radius $\mathcal{R}_i$ corresponding to the distance between the source and the cores in L1688 (see panel 2 of Figure \ref{fig:cartoon} for an illustration of each of these factors); a similar approach as has been taken to infer the sources of $^{60}$Fe deposited on the ocean floor\citep{breitschwerdt_locations_2016}. The resolution of INTEGRAL and COMPTEL are such that any finer structure in the $\gamma$-ray emission is not reliably discernible. The smaller-scale behavior of the $^{26}$Al and its interactions with denser gas in the region encapsulated in $\mathcal{C}$ is the primary uncertainty in this part of the estimate, which we will return to momentarily. Even reasonable values of $\mathcal{R}_i$ are quite uncertain. As one extreme, we adopt a fixed value of $\mathcal{R}_i=10$ pc for all sources. This is reasonable based on the separation between L1688 and the apparent center of the $\gamma$-ray emission visible in Figure 1, but large ambiguity in exactly what value to adopt remains owing to the low significance of the COMPTEL map and the 3D geometry. To approach the question in a principled way, we produce an estimate of the distribution of $\mathcal{R}_i$ values based on the assumption that the present-day massive stars in Upper Sco are drawn from an isotropic 3D gaussian distribution, and that past sources of SLRs follow the same distribution. We derive this distribution taking into account the line of sight distance uncertainties to the present-day massive stars in the methods section, and show the results in Figure 9, which demonstrates that the typical 3D separation between the SLR sources and L1688 is indeed about 10 pc (see the first section of the Supplementary Information). The incorporation of the $^{26}$Al as estimated via the Pre-Enrichment and Accumulation terms is discussed in the second section of the Supplementary Information.

\end{methods}

\begin{addendum}
\item  J.A. acknowledges support from the University of Vienna, in particular the TURIS research platform and the Data-Science research center, and the Radcliffe Institute. J.C.F. acknowledges funding from an ITC Fellowship and a Flatiron Research Fellowship. D.N.C.L. thanks the Institute for Theory and Computation, Harvard University for support while this work was initiated. We thank Stan Woosley, Roland Diehl, Simon Portegies Zwart, Dan Foreman-Mackey, Catherine Zucker, Fran Bartoli\'c, Herv\'e Bouy, and Stefan Meingast for useful conversations, Roland Diehl for providing the COMPTEL $\gamma$-ray map, and Martin Krause and the other (anonymous) referees for their constructive feedback.
 
\item[Author contributions] J.A. provided the observational data used in this Letter and produced Figures 1 and Extended Data Figure 1. J.F. led the forward modelling, and produced the other Figures., D.N.C.L. initiated the collaboration and led the integrated approach. All authors contributed to interpretation of the results and preparation of the manuscript.

\item[Competing interests] The authors declare no competing interests.

\item[Data Availability] Observational data underlying Figures 1, 5, and 6 will be made available upon reasonable request from J.A., with the $\gamma-$ray data subject to approval by R. Diehl. Posterior samples and weights obtained in our inference problem are available as a pickle file at https://github.com/jcforbes/ophiuchus-al26

\item[Code Availability] Code to create the plots and generate posterior samples is available  at https://github.com/jcforbes/ophiuchus-al26.

\item[Correspondence and requests for materials] Correspondence and requests for materials should be addressed to J.F. at jforbes@flatironinstitute.org.

\end{addendum}

{\bf Supplementary Information}

\setlength{\baselineskip}{4ex}


\section{Distance distribution}

One aspect of propagating the $^{26}$Al from the sources, whose modelling we have discussed above in detail, to the cores is knowing the distance between the sources and the cores. The simplest assumption would be to take a fixed distance based on the present-day morphology of the $^{26}$Al. However, given the uncertainties in the fine details of the $\gamma$-ray map, we opt instead to estimate the 3D distribution of possible $^{26}$Al sources in Upper Sco relative to the locations of the present-day cores in L1688. 

\noindent
First, we assume that the massive stars in Upper Sco represent a random draw from some underlying 3D spatial distribution, and that past sources of $^{26}$Al had positions drawn from the same distribution. Next, we combine the locations on the sky of the 21 most massive stars in Upper Sco, including the runaway star $\zeta$ Oph, with distances from Hipparcos, Gaia, and a recent distance catalogue of local molecular clouds$^{8}$. We denote the estimated distances $\hat{d}_i$ and their reported Hipparcos errors $\hat{\sigma}_{d,i}$. We assume negligible errors in the sky locations of the stars, and for simplicity model their underlying 3D distribution as an isotropic gaussian, with unknown 3D center $\vec{r}_c$ and width $w$.

\noindent
The likelihood of observing the collection of massive stars at their present locations is then
\begin{eqnarray}
\label{eq:distlik}
    \mathcal{L}(\{\alpha_i\}, \{\delta_i\} | \vec{r}_c, w, \{d_i\}) =
    \prod_{i=1}^{21} \left[ \frac{1}{\sqrt{2\pi w^2}}\exp{\left( \frac{-(\vec{r}_i - \vec{r}_c)^2}{2 w^2} \right)}  \right]  
\end{eqnarray}
Quantities in brackets $\{x_i \}$ refer to properties of the individual stars, indexed by $i$. These are the observed right ascension and declination, $\alpha_i$ and $\delta_i$, and the (unknown) true distances $d_i$. The position of the $i$th star in 3D is $\vec{r}_i$, which is set by $\alpha_i$, $\delta_i$, and $d_i$.

\noindent
In order to incorporate our information about the distance to each star, we can treat the estimated distances $\hat{d}_i$ and their reported uncertainties $\hat{\sigma}_{d,i}$ as a Gaussian prior on the true distances $d_i$. We do so with one modification, namely that the reported Hipparcos uncertainties are multiplied by a constant $f_e$ which is the same for all stars. Applying Bayes' Theorem,
\begin{eqnarray}
\label{eq:distpost}
p(\{d_i\}, f_e, \vec{r}_c, w | \{\hat{d}_i\}, \{\hat{\sigma}_{d,i}\}, \{\alpha_i\}, \{\delta_i\}) &\propto& p(f_e, \vec{r}_c, w) \left( \prod_{i=1}^{21} p(d_i | f_e,  \{\hat{d}_i\}, \{\hat{\sigma}_{d,i}\}) \right) \nonumber \\
& & \ \ \times \mathcal{L}(\{\alpha_i\},\{\delta_i\}|\vec{r}_c,w,d_i),
\end{eqnarray}
where the prior distribution over the true distances is just taken to be a normal distribution,
\begin{equation}
   p(d_i | f_e,  \{\hat{d}_i\}, \{\hat{\sigma}_{d,i}\}) =  \frac{1}{\sqrt{2\pi f_e^2 \hat{\sigma}_{d,i}^2}}\exp{\left( \frac{-(d_i - \hat{d}_i)^2}{2 f_e^2 \hat{\sigma}_{d,i}^2} \right)}.
\end{equation}
The advantage of writing the problem in this way is that, for our purposes, the true $d_i$ are nuisance parameters that we can now marginalize out. This would ordinarily require a 21-dimensional integral over all $d_i$, but we can see from equations \ref{eq:distlik} and \ref{eq:distpost} that each integral over $d_i$ can be performed separately, so that we are left with a marginalized posterior,
\begin{equation}
    p( f_e, \vec{r}_c, w | \{\hat{d}_i\}, \{\hat{\sigma}_{d,i}\}, \{\alpha_i\}, \{\delta_i\}) \propto p(f_e, \vec{r}_c, w) \prod_{i=1}^{21} \int_0^\infty p(d_i|f_e,  \{\hat{d}_i\}, \{\hat{\sigma}_{d,i}\}) \mathcal{L}_i (\alpha_i,\delta_i|\vec{r}_c,w,d_i) \mathrm{d}d_i
\end{equation}
where $\mathcal{L}_i$ refers to the $i$th factor in the likelihood given by equation \eqref{eq:distlik}. These integrals may be thought of as weighted sums along the line of sight through the assumed underlying 3D distribution. While they are not analytic, they are 1-dimensional, making them quite tractable and transforming the inference problem from 26-dimensional down to 5-dimensional.

\noindent
We then sample from the posterior distribution with an affine-invariant ensemble sampler$^{81}$. Our prior is that $f_e$ is log-normal, with a median of 1 and a standard deviation in the log of 0.3 dex. The distance to the center of the putative 3D distribution is {\it a priori} uniform in the log from 10 pc to 1000 pc, and the size of the 3D distribution is uniform in the log from 0.1 pc to 100 pc. To ensure a uniform prior on the celestial sphere for the location of $r_c$, namely $\alpha_c$ and $\delta_c$, we impose a flat prior on $\alpha_c$ and the cosine of the angle from the north celestial pole $\cos(\pi/2 -  \delta_c)$.

\noindent
While the full posterior distribution may be of some interest, for our purposes we want a single instance of the model from which to draw 3D locations of SLR sources. A natural choice for which instance of the model to use is the maximum {\it a posteriori} model. The resulting distribution of distances between the SLR sources and L1688 is shown in Extended Data Fig. 7. In practice we use the best estimate of the 3D location of Elia 2-29 as a proxy for the location of L1688 to produce the distance distribution in this figure. The model 3D distribution shown here is employed in computing the salmon ``Broad Distance Distribution'' lines in Figure 4.

\section{Accumulation}

For a particular value of $\rho_{^{26}\mathrm{Al}}$, we estimate the values of the Pre-Enrichment and Accumulation terms as follows. In the Pre-Enrichment term, the $^{26}$Al mass is simply the volume of the material that will eventually form the core times the density of $^{26}$Al at the time the core formed, times the appropriate exponential decay factor,
\begin{equation}
    M_{^{26}\mathrm{Al}, \mathrm{core}, \mathrm{pre}} = \frac43 \pi r_c^3 \chi \rho_{^{26}\mathrm{Al}}(t = t_\mathrm{obs} - t_c) 2^{-t_c/\tau_{1/2}}.
\end{equation}
Here $\chi>1$ is the density contrast between the cores and the lower-density medium from which they form. We adopt $\chi=100$ and $r_c=8000$ AU. The latter is consistent with the cores shown in Extended Data Figure 1, though there is some dispersion. We can estimate the density contrast by noting that typical dust column densities in L1688 are about a factor of 10 lower than the columns through the cores. If we assume that the spatial extent of the cores, and of L1688 itself, is comparable to its size on the plane of the sky, we would conclude that the volume density contrast is just the column density contrast times the ratio of the objects' sizes. For L1688 and its cores, we have $0.5 \mathrm{pc}/r_c \approx 13$, so $\chi=100$ is reasonable, though the true value may vary by factors of several given the uncertainties and the physical variation between the different cores.

\noindent
Meanwhile the accumulation term requires summing up the contributions from material flowing past the core over the course of its lifetime, 
\begin{equation}
\label{eq:accum}
    M_{^{26}\mathrm{Al}, \mathrm{core}, \mathrm{accum}} = \pi r_c^2 v_\mathrm{rel} \epsilon  \int_{t_{\mathrm{obs}}-t_c} ^{t_\mathrm{obs}} 
    \rho_{^{26}\mathrm{Al}}(t') 2^{-( t_\mathrm{obs} - t')/\tau_{1/2}} dt'.
\end{equation}
We include an efficiency factor $\epsilon$ since only a fraction of the material that intersects the core's path through the inter-core medium is actually accreted onto the core. Several arguments suggest that $\epsilon \sim 1\%$. First, idealized simulations of shockwaves triggering the collapse of cores$^{13,82}$ measure this quantity directly, and find values of this order depending on the state of the gas, especially the Mach number$^{83}$. The ratio of the cross-sectional area described by the Bondi radius $\propto r_\mathrm{Bondi}^2 = G^2 M_c^2/v_\mathrm{rel}^4$ to the physical cross-section of the core $\propto r_c^2$ also turns out to be $\sim 1\%$ for a solar mass core of size 8000 AU and  relative velocity 1 km s$^{-1}$. While there is substantial uncertainty in this parameter, it will turn out to be largely irrelevant as long as $\epsilon \ll 1$. The relative velocity between the core and the lower-density material, $v_\mathrm{rel}$ may be estimated observationally by examining the line-of-sight velocity of the gas in the region as traced by $^{13}$CO and NH$_3$. Individual cores traced by the latter are seen to have distinct velocities offset from the surrounding $^{13}$CO by a few km/s (see Figure \ref{fig:velocity}).

\noindent
We now turn to the value of the concentration parameter $\mathcal{C}$. The simplest approach would be to set $\mathcal{C}\approx 1$. This assumption is reasonable in the low-density high-temperature gas within Upper Sco, but is less plausible as the $^{26}$Al-bearing material mixes with the denser gas of L1688 and its surroundings. Simulations of mixing in the interstellar medium provide some guidance. First, zoom-in simulations$^{19}$ beginning from an isolated galaxy simulation$^{24}$ demonstrate that during the collapse of structure in the molecular clouds, metal inhomogeneities on scales $\lesssim 1$ pc are efficiently mixed together, suggesting that as the $^{26}$Al-bearing gas interacts with L1688 and the gas from which it formed, the $^{26}$Al available (as determined by the constant-density sphere of radius $\mathcal{R}_i$) would not have great difficulty mixing with the denser gas such that $\mathcal{C}$ really may be described by a constant. In the space of density vs. $^{26}$Al/$^{27}$Al, i.e. concentration, simulations of a turbulent periodic box with sink particles and SLR injection by supernovae$^{23}$ showed$^{18}$ (their Fig. 2) that the maximum SLR concentration for a given density scales approximately as $\rho^{-1/3}$ interspersed with a constant scaling as individual objects collapse and cease to mix. In contrast, if SLR-free gas is mixed with a fixed volume of SLR-bearing gas as in the $\mathcal{C}=1$ case, the concentration would scale as $\rho^{-1}$. The dust column density in the unambiguously $^{26}$Al-bearing region is roughly 100 times lower than the column density within L1688.
Given the difference in physical scale on the sky of these two regions, about a factor of 6, the volume density contrast is likely $\lesssim 10^3$. One can consider $\mathcal{C}$ to be a factor accounting for the difference between $^{26}$Al/$^{27}$Al $\propto \rho^{-1}$ and the true scaling. If we adopt $\rho^{-1/3}$ as the true scaling, we obtain $\mathcal{C}\propto \rho^{2/3}$, so for L1688, $\mathcal{C} \sim 100$, which we adopt going forward.

\noindent
Adopting fiducial values for $\chi, r_c, v_\mathrm{rel}$, $\mathcal{C}$ and $\epsilon$, and assuming each has an independent factor of 2 uncertainty owing to geometric and line-of-sight uncertainties, we use our ensemble of $M_{^{26}\mathrm{Al}}(t)$ histories of Upper Sco to compute the median and central 68\% of the values of $M_{^{26}\mathrm{Al}, \mathrm{core}, \mathrm{accum}}$ and $ M_{^{26}\mathrm{Al}, \mathrm{core}, \mathrm{pre}}$ under the two different assumed source distance $\mathcal{R}_i$ distributions (fixed, and broad, corresponding to a reasonable estimate of the 3D distribution of massive stars in Upper Sco). Figure 4 shows the results for a varying value of $t_c$, along with an errorbar representing an observational estimate and its quoted factor-of-two uncertainty$^{17}$ of $t_c$ in this region. Note that the shaded regions around each line represent the uncertainty from both the $^{26}$Al production model and the assumed factor of 2 uncertainties in the fiducial parameters determining how the $^{26}$Al is incorporated into cores. In the case of the ``Broad Distance Distribution,'' it also includes dispersion from the spread in distances $\mathcal{R}$ between the sources of SLRs and L1688. 

\renewcommand\thefigure{\arabic{figure}}    
\setcounter{figure}{0}

\begin{figure}
\includegraphics[width=1\linewidth]{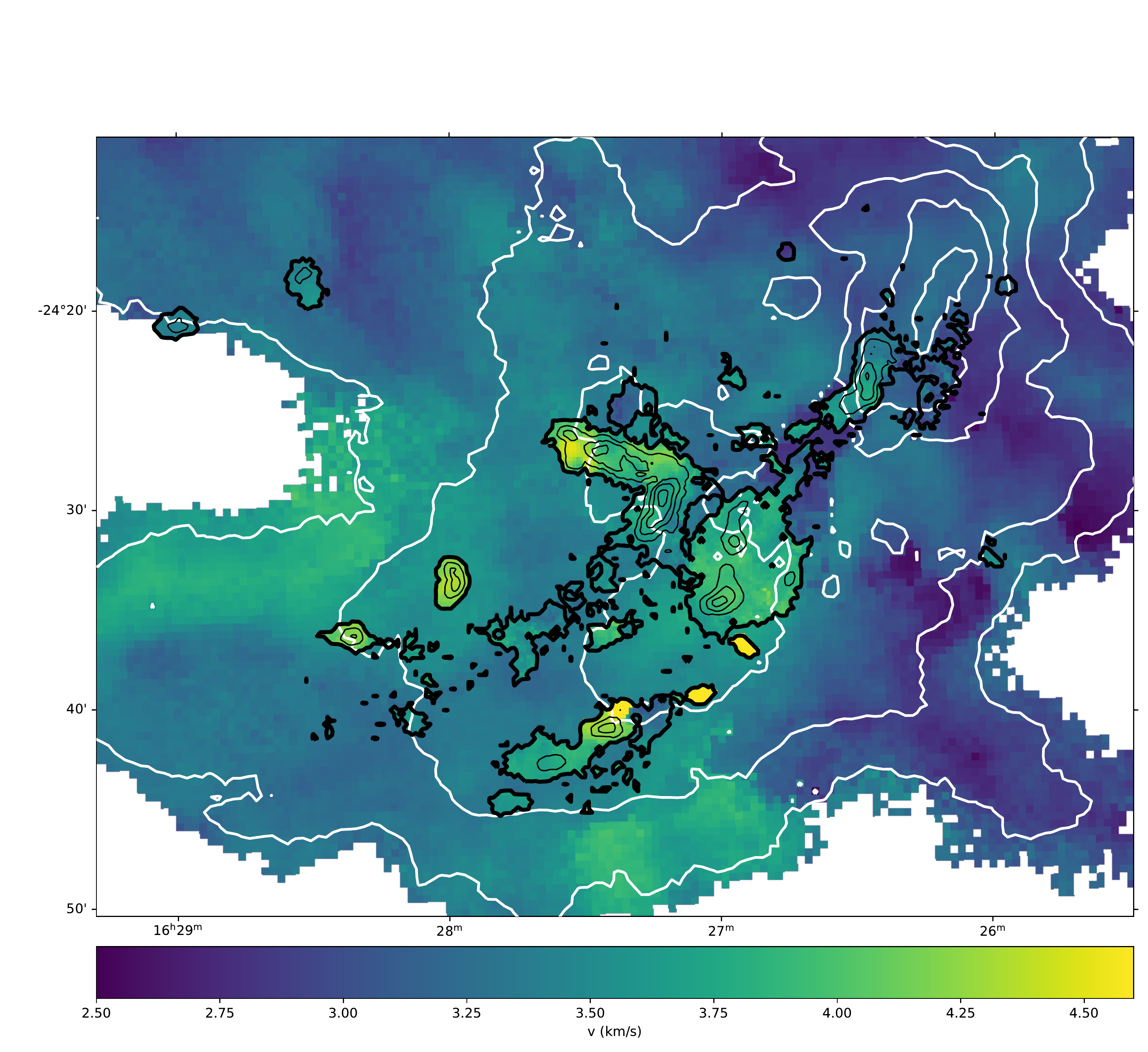} 
\caption{[Supplementary Information] Relative motion of the cores and background gas. L1688 is shown here in a combination of $^{13}$CO from the GAS survey$^{51}$ and NH$_3$ from the COMPLETE survey$^{52}$. For each tracer, we compute the mean velocity map. In regions where data from both tracers are available, we show only the NH$_3$. Black contours show the intensity of NH$_3$, and white contours show the intensity of $^{13}$CO. Outside of the NH$_3$, i.e. outside of the thick black contour line, the color shows the mean velocity of the $^{13}$CO.  Cores, clearly visible in the NH$_3$, tend to have 1-2 km s$^{-1}$ offsets in mean line-of-sight velocity from the surrounding gas.}
\label{fig:velocity}
\end{figure}

\begin{figure}
\includegraphics[width=1\linewidth]{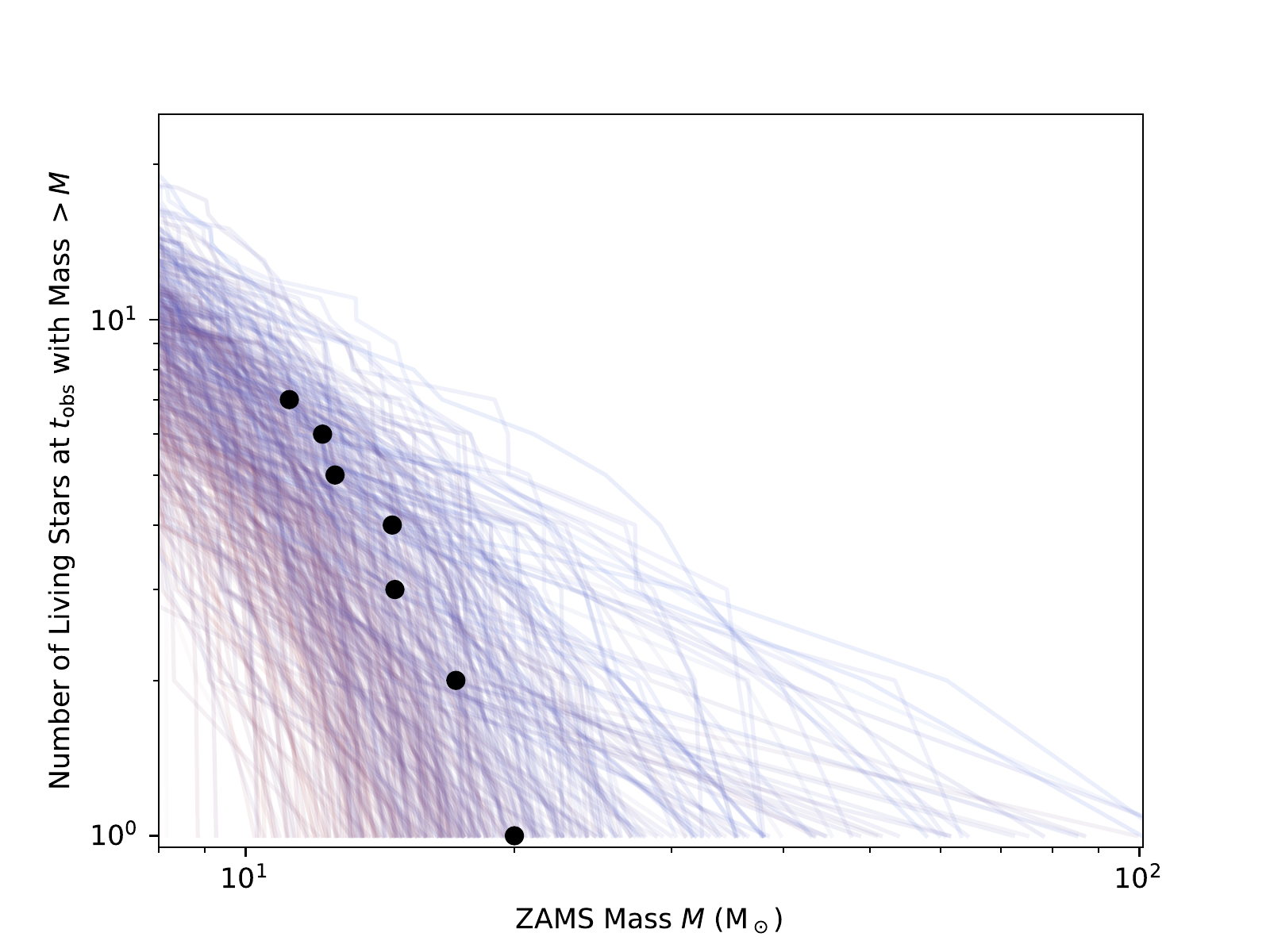} 
\caption{[Supplementary Information] The predicted distribution of stars alive today in Upper-Sco. The black dots show estimates$^{46}$ for the 6 most massive stars in Upper-Sco, plus $\zeta$ Oph\citep{}, which is a runaway star from Upper-Sco. Lines are colored by the age of the association in that realization of the model with the reddest colors corresponding to the oldest ages. The fact that the points lie in the most likely part of the distribution provides some validation of our model, and demonstrates that in the future using information about the mass of currently-living stars, in addition to the mass of living $^{26}$Al, could substantially narrow down the range of possible scenarios for the evolution of Upper-Sco.}
\label{fig:cdf}
\end{figure}

\end{document}